\documentclass[prx,twocolumn,showpacs,floatfix,preprintnumbers,amsmath,amssymb,superscriptaddress]{revtex4}

\usepackage{graphicx}
\usepackage{amsmath}
\usepackage{amssymb}
\usepackage{color}
\usepackage{dcolumn}
\usepackage{epsfig}
\usepackage{bm}
\usepackage[urlcolor=blue]{hyperref}
\hypersetup{backref, colorlinks=true, linkcolor=blue, citecolor=blue}

\begin{document}

\preprint{}

\title{ {NbS$_{3}$: A unique quasi one-dimensional conductor with three charge density wave transitions}}

\author{S.G. Zybtsev}
\affiliation{Kotel$'$nikov Institute of Radioengineering and Electronics of RAS, Mokhovaya 11-7, 125009 Moscow, Russia}

\author{V.Ya. Pokrovskii}
\email{pok@cplire.ru}
\affiliation{Kotel$'$nikov Institute of Radioengineering and Electronics of RAS, Mokhovaya 11-7, 125009 Moscow, Russia}

\author{V.F. Nasretdinova}
\affiliation{Kotel$'$nikov Institute of Radioengineering and Electronics of RAS, Mokhovaya 11-7, 125009 Moscow, Russia}

\author{S.V. Zaitsev-Zotov}
\affiliation{Kotel$'$nikov Institute of Radioengineering and Electronics of RAS, Mokhovaya 11-7, 125009 Moscow, Russia}

\author{V.V. Pavlovskiy}
\affiliation{Kotel$'$nikov Institute of Radioengineering and Electronics of RAS, Mokhovaya 11-7, 125009 Moscow, Russia}

\author{A.B. Odobesco}
\affiliation{Kotel$'$nikov Institute of Radioengineering and Electronics of RAS, Mokhovaya 11-7, 125009 Moscow, Russia}

\author{Woei Wu Pai}
\email{wpai@ntu.edu.tw}
\affiliation{Center for condensed matter sciences, National Taiwan University, Taipei, Taiwan, 106}

\author{M.-W. Chu}
\affiliation{Center for condensed matter sciences, National Taiwan University, Taipei, Taiwan, 106}

\author{Y. G. Lin}
\affiliation{National synchrotron research center, HsinChu, Taiwan 300}

\author{E. Zupani\v{c}}
\affiliation{Jo\v{z}ef Stefan Institute, Jamova 39, Ljubljana, Slovenia}

\author{H.J.P. van Midden}
\affiliation{Jo\v{z}ef Stefan Institute, Jamova 39, Ljubljana, Slovenia}

\author{S. \v{S}turm}
\affiliation{Jo\v{z}ef Stefan Institute, Jamova 39, Ljubljana, Slovenia}

\author{E. Tchernychova}
\affiliation{Jo\v{z}ef Stefan Institute, Jamova 39, Ljubljana, Slovenia}

\author{A. Prodan}
\affiliation{Jo\v{z}ef Stefan Institute, Jamova 39, Ljubljana, Slovenia}

\author{J.C. Bennett}
\affiliation{Department of Physics, Acadia University, Wolfville, N. S. Canada}

\author{I.R. Mukhamedshin}
\affiliation{Institute of Physics, Kazan Federal University, 420008 Kazan, Russia}

\author{O.V. Chernysheva}
\affiliation{National Research Nuclear University "MEPhI" (Moscow Engineering Physics Institute), Kashirskoe sh. 31, 115409 Moscow, Russia}

\author{A.P. Menushenkov}
\affiliation{National Research Nuclear University "MEPhI" (Moscow Engineering Physics Institute), Kashirskoe sh. 31, 115409 Moscow, Russia}

\author{V.B. Loginov}
\affiliation{National Research Nuclear University "MEPhI" (Moscow Engineering Physics Institute), Kashirskoe sh. 31, 115409 Moscow, Russia}

\author{B.A. Loginov}
\affiliation{National Research University of Electronic Technology (MIET), 124498, Zelenograd, Moscow, Russia}

\author{A.N.Titov}
\affiliation{Ural Federal University, Mira 19, 620002, Yekaterinburg, Russia}

\author{Mahmoud Abdel-Hafiez}
\email{m.mohamed@hpstar.ac.cn}
\affiliation{Center for High Pressure Science and Technology Advanced Research, Beijing 100094, China}
\date{\today}

\begin{abstract}

Through transport, compositional and structural studies, we review the features of the charge-density wave (CDW) conductor of NbS$_{3}$ (phase II).  We highlight three central results: 1) In addition to the previously reported CDW transitions at $T_{P1}$ = 360\,K and $T_{P2}$ = 150\,K, another CDW transition occurs at a much higher temperature $T_{P0}$ = 620-650\,K; evidence for the non-linear conductivity of this CDW is presented. 2) We show that CDW associated with the $T_{P2}$ - transition arises from S vacancies acting as donors. Such a CDW transition has not been observed before. 3) We show exceptional coherence of the $T_{P1}$-CDW at room-temperature. Additionally, we report on the effects of uniaxial strain on the CDW transition temperatures and transport.

\end{abstract}

\pacs{Condensed Matter Physics, Materials Science, Superconductivity}
\maketitle

\section{Introduction}

Since Peierls transitions, at which electrons condense into charge-density waves (CDWs), usually occur well below room temperature (RT), studies of CDWs in quasi one-dimensional (1D) conductors have been usually considered a branch of low-temperature physics~\cite{1,2}. The formation of a CDW is accompanied by dielectrization (i.e., gapping) of the electronic spectrum with a corresponding drop in electrical conductivity. Periodic lattice distortion accompanied with the CDW can be studied with both diffraction techniques in momentum space and scanning tunneling microscopy in real space. A notable basic feature of quasi 1D CDWs is their ability to slide in a sufficiently high electric field, resulting in non-linear conductivity. This sliding is accompanied by the generation of narrow- and wide-band noises. The quasi 1D CDWs also feature an enormous dielectric constant and metastable states originating from their deformability. In addition to these basic effects, a number of other features have been investigated including synchronization of CDW sliding with an external radio-frequency (RF) field (the so-called Shapiro steps), coherence stimulation of CDW sliding by asynchronous RF irradiation~\cite{3}, the effects of pressure~\cite{2} and uniaxial strain~\cite{4,5,6,7} on the Peierls transitions and CDW transport, and enormous electric-field-induced crystal deformations~\cite{2,8}.

Several trichalcogenides of the group V metals (MX$_{3}$), namely NbSe$_{3}$, TaS$_{3}$ (orthorhombic and monoclinic) and phase II NbS$_{3}$ (hereafter NbS$_{3}$-II), constitute a family of typical quasi 1D CDW conductors~\cite{1,2}. Their crystal structures are formed from metallic chains surrounded by trigonal prismatic cages of chalcogen atoms. Though these compounds are apparently isoelectronic, their properties are rather diverse. For example, they display very different CDW wave vector magnitudes, indicating different degrees of filling in conduction electronic bands. A plausible reason for this variety may be the relative positioning of the chalcogen atoms~\cite{9}. Depending on their interatomic distances, chalcogen atoms can either be isolated from each other or they form bonded pairs. Correspondingly, one valence electron from a chalcogen atom can either belong to the conduction band or to a localized bond.

The monoclinic polymorph NbS$_{3}$-II exhibits some fascinating CDW features. Two CDW transitions have been reported for NbS$_{3}$-II. One CDW has a wave vector $q_{1}$ = (0.5$a$*, 0.298$b$*, 0)~\cite{10,11}, and forms at $T_{P1}$=330-370 K~\cite{11,12}, a temperature much beyond the traditional realm of low-temperature physics. While several basic experimental results on NbS$_{3}$-II were published in the 1980s ~\cite{1,2,10,11,13,14,15,155}, little subsequent work was undertaken until 2009. This gap was largely due to difficulties with synthesizing the compound: NbS$_{3}$-II whiskers were only found as rare inclusions accompanying the growth of the semiconducting phase of NbS$_{3}$ (hereafter NbS$_{3}$-I)~\cite{11}, which was more extensively studied~\cite{1,2,16,17}. A new phase-III NbS$_{3}$ with a phase transition at 150\,K was also reported~\cite{13}, which was later suggested to be a sub-phase of NbS$_{3}$-II~\cite{8,12,18}. The synthesis conditions of NbS$_{3}$-II were established from studies in the Kotel'nikov Institute of Radioengineering and Electronics of RAS in 2009~\cite{8}, and were recently successfully reproduced in the National Taiwan University. A detailed description of NbS$_{3}$-II growth conditions is presented in~\cite{8}.

The most notable feature of NbS$_{3}$-II is its non-linear transport at RT~\cite{1,2,11,12,18} associated with the CDW formed at $T_{P1}$. The RT-CDW shows an exceptionally high transport coherence with the highest reported transport velocities of any known sliding CDW. The corresponding values of the fundamental frequency, $f_{f}$, as revealed by the RF interference technique~\cite{12,18}, exceed 15\,GHz. The coherence of this CDW can be further improved by external asynchronous RF irradiation~\cite{3} and by uniaxial strain~\cite{19}. The Peierls transition at $T_{P1}$ is clearly detected in transport and structural studies: a pronounced increase of resistance, $R$, with decreasing temperature is observed near $T_{P1}$, while its I-V curve becomes nearly linear above 340\,K. The intensities of the $q_{1}$ satellites strongly decrease significantly above 350\,K~\cite{11,20}, while the satellites of second modulation wave vector $q_{0}$ = (0.5$a$*, 0.352$b$*, 0) remain detectable to at least 450\,K.

Within NbS$_{3}$-II two "sub-phases" have been identified: a low-ohmic and a high-ohmic sub-phase~\cite{8,18,21}. According to~\cite{8}, the low-ohmic samples are prepared at 670-700\,$^{\circ}$C, and the high-ohmic ones at 715-740\,$^{\circ}$C. Further experiments have shown that low-ohmic samples can also be synthesized between 720 and 730\,K in the presence of at. 16\% excess of S. Both sub-phases can grow in the same run if a temperature gradient is present within the synthesis ampoule. Electron diffraction reveals a doubling of lattice constant along the $a$ axis for the high-ohmic crystals. Such a lattice doubling is absent in the low-ohmic ones~\cite{18}. In addition to the CDW transition at $T_{P1}$, the low-ohmic samples undergo a further CDW transition at $T_{P2}$ = 150\,K~\cite{13} as detected in the temperature-dependent resistance, $R(T)$, curves~\cite{8,12,18}. Below $T_{P2}$, a non-linear conduction with a pronounced threshold field, $E_{t}$, is observed. This indicates a charge transport coupled to this low-temperature CDW (LT CDW). The presence of Shapiro steps provides definitive proof of LT CDW sliding~\cite{12,18}. RF synchronization studies reveal a surprisingly low charge density of this LT-CDW. The so-called "fundamental ratio", $j_{c}$/$f_{f}$, appears very low and sample dependent~\cite{12,18} (here $j_{c}$ is the CDW current density at the 1$^{st}$ Shapiro step). Electron and X-ray diffraction studies indicated that the low-ohmic samples were homogeneous rather than a mixture of phases. Several attempts to find structural changes below $T_{P2}$ were unsuccessful~\cite{10,22}. The LT-CDW state emerges from a dielectrized state following the two CDWs formed at higher temperatures. Therefore, this CDW remains a rather enigmatic charge-ordered state. The emergence of a new CDW in this rather resistive state is very unusual and a Keldysh-Kopaev transition~\cite{23} (the formation of an excitonic dielectric~\cite{24}) has been proposed as a possible mechanism~\cite{18}.

In this paper we present a number of new experimental results for NbS$_{3}$-II. Section \emph{II} focuses on the features of the RT-CDW. We report unprecedentedly high values of CDW fundamental frequencies as revealed by Shapiro steps. A high coherence of the RT-CDW sliding is shown by the nearly complete CDW synchronization under RF power and by Bessel-type oscillations of the Shapiro steps' width vs. RF power. In addition, we experimentally demonstrate that the coherence can be further improved by applying strain, $\epsilon$, parallel to the chains, i.e., along the $b$ axis. Such strain also strongly affects the CDW transition temperature $T_{P1}$ which decreases by approximately 80\,K for $\epsilon$ $\sim$ 1.5\%. In section \emph{III} we report transport measurements at temperatures up to about 650\,K. A new feature in $R(T)$ is found near 620-650\,K and is attributed to the onset of an the ultra-high-temperature (UHT-) CDW. Evidence for non-linear transport provided by the UHT-CDW is given. We also demonstrate that by heating above $\approx$ 800\,K the high-ohmic sub-phase transforms gradually into the low-ohmic sub-phase and further into a metallic-like compound. Section \emph{IV} illustrates and focuses on the LT-CDW. Unlike the transition at $T_{P1}$, the LT-CDW transition appears nearly insensitive to strain, as is the non-linear conduction associated with this LT-CDW. Electron-probe microanalyses (EPMA) reveal a shortage of S in the low-ohmic samples. This suggests a coupling between the relatively high conductivity of the low-ohmic samples and the presence of S vacancies. Though electron diffraction patterns show no changes below 150\,K, the $T_{P2}$ transition is detected by TEM-based electron energy loss spectroscopy (STEM-EELS) and X-ray absorption near edge spectroscopy (XANES). These techniques indicate charge transfers between states coupled with S and Nb atoms, as well as with S vacancies. Nuclear magnetic resonance (NMR) studies suggest "freezing" of the condensed electronic state near $T_{P2}$. In section \emph{V}, we present an overview of the results and discuss various possible origins of the LT-CDW transition. The Keldysh-Kopaev (excitonic insulator) transition~\cite{23,24} appears more consistent with the experimental data then other mechanisms of electronic condensation. In the concluding section \emph{VI}, we summarize the main features of NbS$_{3}$-II.

\section{ROOM-TEMPERATURE CDW; EFFECTS OF RF IRRADIATION AND UNIAXIAL STRAIN}

The RT-CDW in NbS$_{3}$-II has been most thoroughly studied to date. It is remarkable not only for the occurrence of sliding, but also for its extremely high sliding coherence. The "fundamental ratio", $j_{c}$/$f_{f}$, is sample independent within the experimental error (defined by the uncertainty in the sample cross-sectional area). The experimentally determined value $j_{c}$/$f_{f}$=18 A/MHz/cm$^{2}$~\cite{18} corresponds to one CDW chain per unit cell~\cite{12}. The universality of this "fundamental ratio" allows a precise determination of the cross-section area of the samples~\cite{25}. A very high coherence of the RT-CDW sliding is observed in samples with nanoscale transverse dimensions. In these samples the highest fundamental frequencies are achieved. As Fig.\,1 shows, Shapiro steps (the first harmonic) are observed at frequencies as high as 20\,GHz. The large area per CDW chain, $s_{0}$=2$e/(j_{c}$/$f_{f}$) = 180 {\AA}$^{2}$~\cite{26}, compared to other quasi-1D CDW compounds, is consistent with the relatively small amount of Joule heating of the NbS$_{3}$-II samples. They show the highest density of CDW current before burn-out~\cite{18}. Based on the highest current densities ever passed through these samples, it is suggested~\cite{18} that fundamental frequencies as high as $f_{f}$ = 200\,GHz are attainable. However, synchronization of the CDW at frequencies above 20\,GHz requires special arrangement of the samples for a better impedance match with RF radiation.

\begin{figure}[tbp]
\includegraphics[width=21pc,clip]{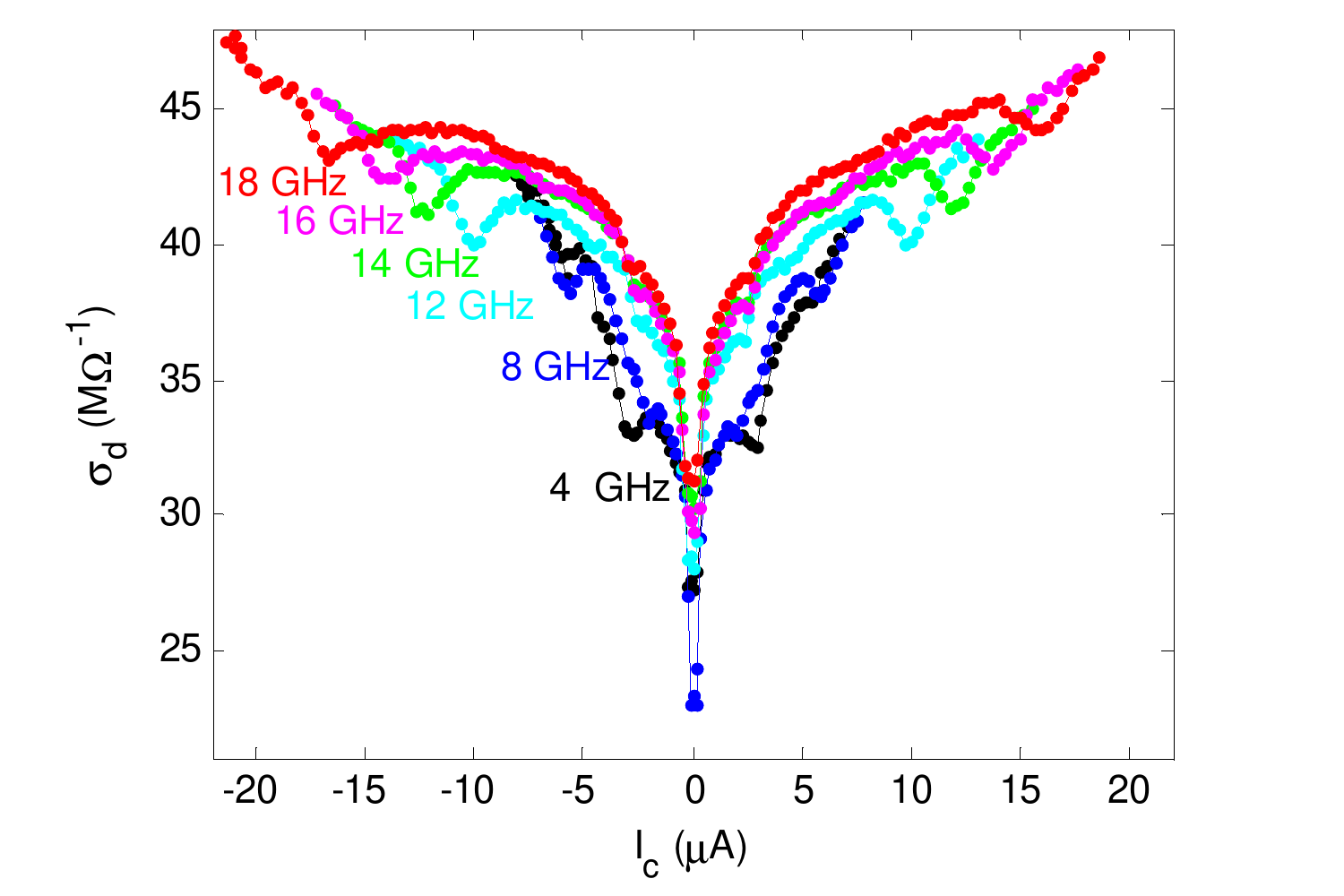}
\caption{The RT dependences of differential conductivity, $\sigma _{d}$, vs. non-linear current, $I_{c}$, under RF irradiation at different frequencies. The sample dimensions are 1.6$\mu$m $\times$ 0.005$\mu$m$^{2}$.} \label{PD}
\end{figure}

At a sufficiently large RF power, the RT-CDW in NbS$_{3}$-II nano-dimensional samples show nearly complete synchronization: the differential conductivity, $\sigma _{d}$, of the CDW is reduced by 80-90\% (Fig.\,2). At the same time, the widths of the Shapiro steps show a non-monotonic dependence on the RF power with Bessel-type oscillations. To observe such oscillations, the CDW should exhibit a high coherence. Previously, the Bessel-type oscillations were observed for NbSe$_{3}$ only~\cite{28}. The evolution of the I-V curves with increasing RF power is shown by the video-clip in the Supplementary Material. It shows the screen of a digital oscilloscope displaying the rapidly recorded I-V curves of a NbS$_{3}$-II sample. Figure 3 presents the normalized dependence of the 1$^{st}$ Shapiro step width and the threshold voltage, $V_{t}$, (half-width of the "0-th step") on the RF power. A non-monotonic suppression of $V_{t}$ has been previously reported for NbSe$_{3}$~\cite{28,29}.

\begin{figure}[tbp]
\includegraphics[width=21pc,clip]{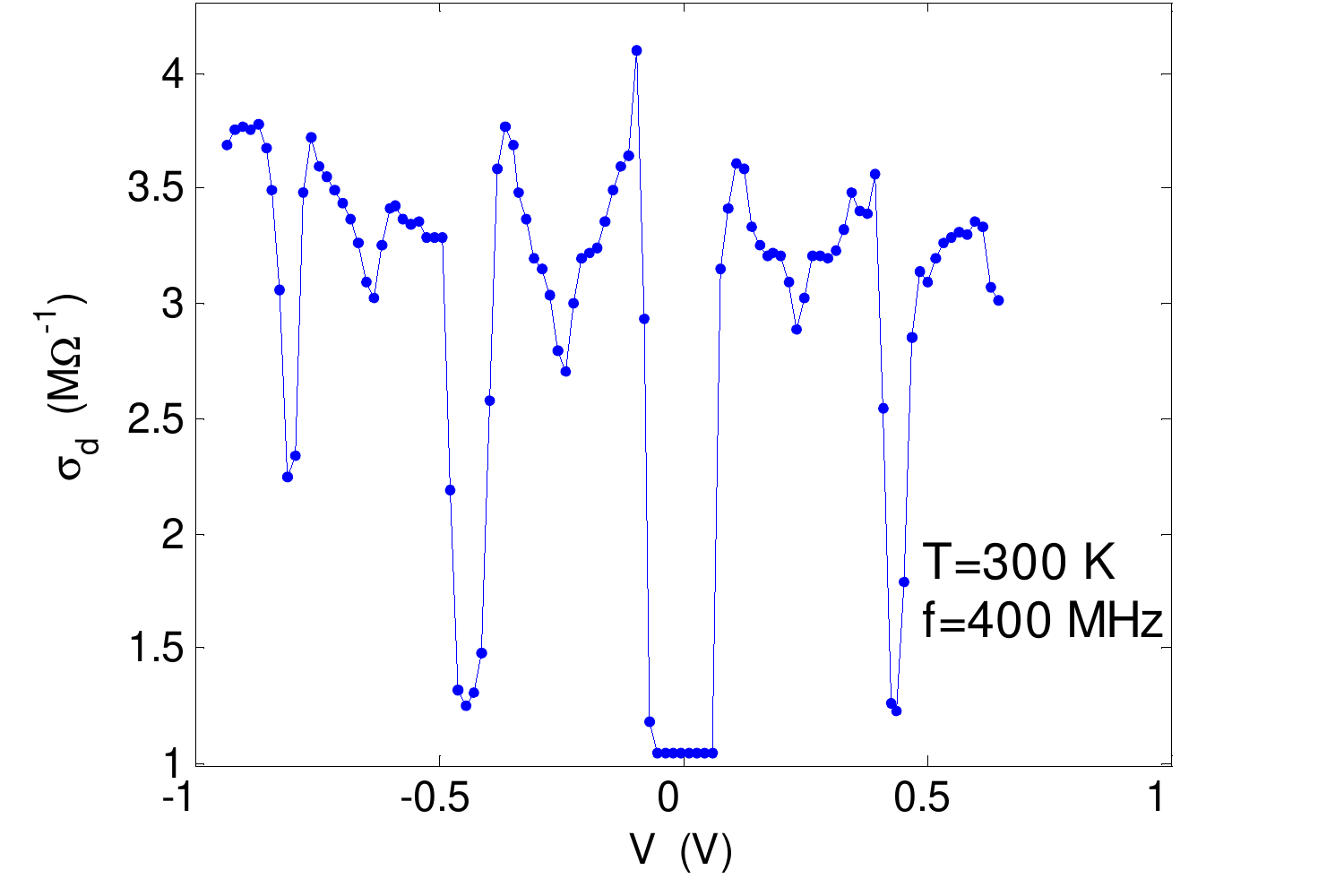}
\caption{An example of $\sigma _{d}$ vs. V dependence under RF irradiation at RT. The sample dimensions are 50$\mu$m $\times$ 0.011$\mu$m$^{2}$.} \label{PD}
\end{figure}

We recently developed techniques to apply controlled uniaxial strain to whisker-like samples, including nano-sized ones~\cite{18}. The NbS$_{3}$-II samples were stretched by bending epoxy-based substrates~\cite{18}. Studies of NbS$_{3}$-II at RT~\cite{19} and of orthorhombic TaS$_{3}$~\cite{19,30} have demonstrated that strain can improve the CDW coherence. Figure 4 summarizes a set of curves showing the differential conductivity of NbS$_{3}$-II as a function of voltage, $V$. Unlike in case of the orthorhombic TaS$_{3}$, in which an ultra-coherent CDW emerges through a phase transition at a critical strain~\cite{19,30}, CDW coherence in case of NbS$_{3}$-II grows gradually with $\epsilon$ (Fig.\,4): the thresholds become sharper, the growth of $\sigma _{d}$ above the thresholds occurs faster, and the value of the maximum CDW conductivity increases~\cite{19}. After strain removal, the threshold voltage decreases and the threshold becomes slightly sharper, while the value of resistance at $V$ = 0 indicates that the sample is unstrained.

\begin{figure}[tbp]
\includegraphics[width=21pc,clip]{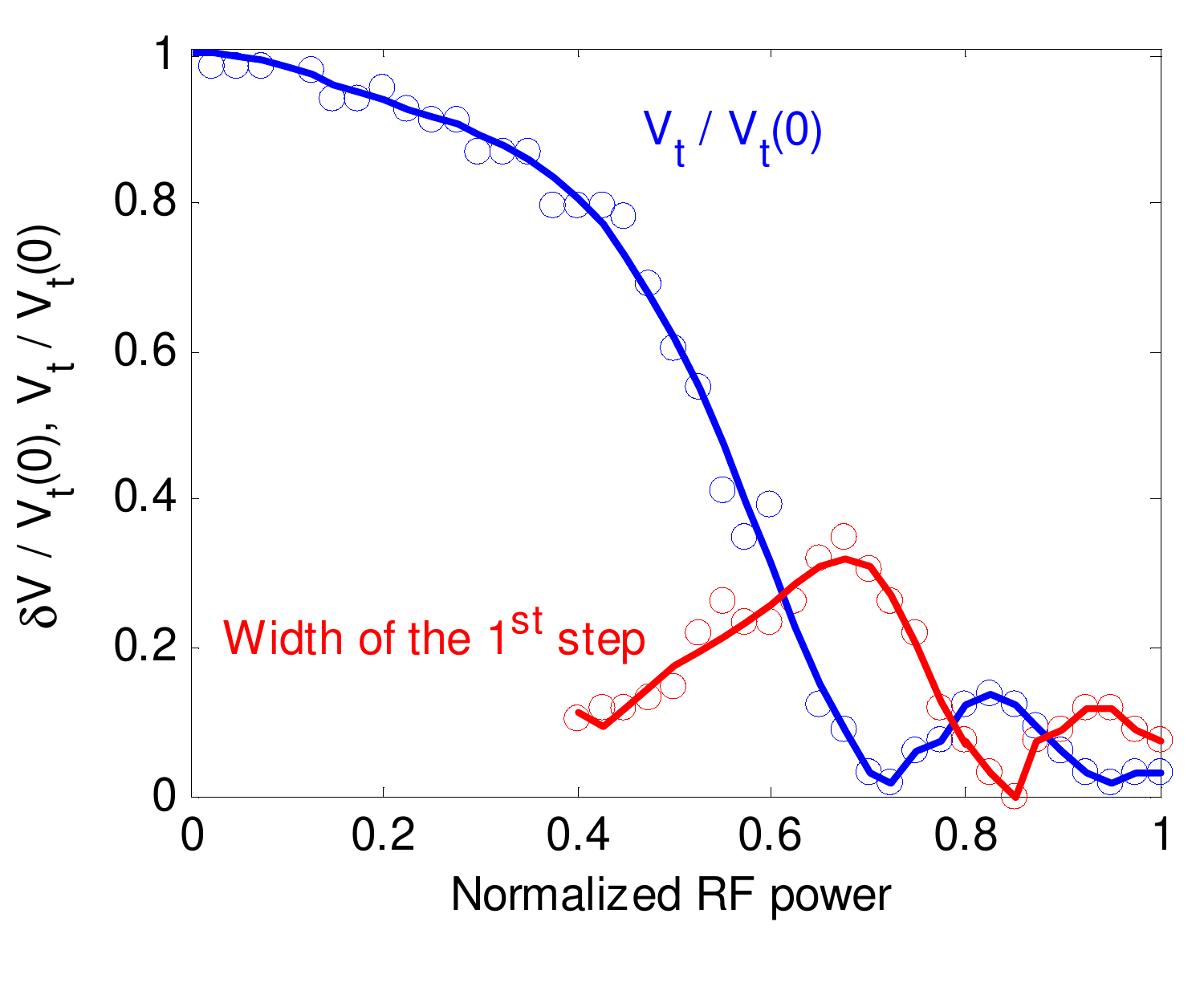}
\caption{A plot of a normalized Shapiro step width and $V_{t}$ vs. RF power (400 MHz). The sample dimensions are 50$\mu$m $\times$ 0.011$\mu$m$^{2}$ (see Fig. 2). The original I-V curves are in the Supplementary Material.} \label{PD}
\end{figure}

\begin{figure}[tbp]
\includegraphics[width=21pc,clip]{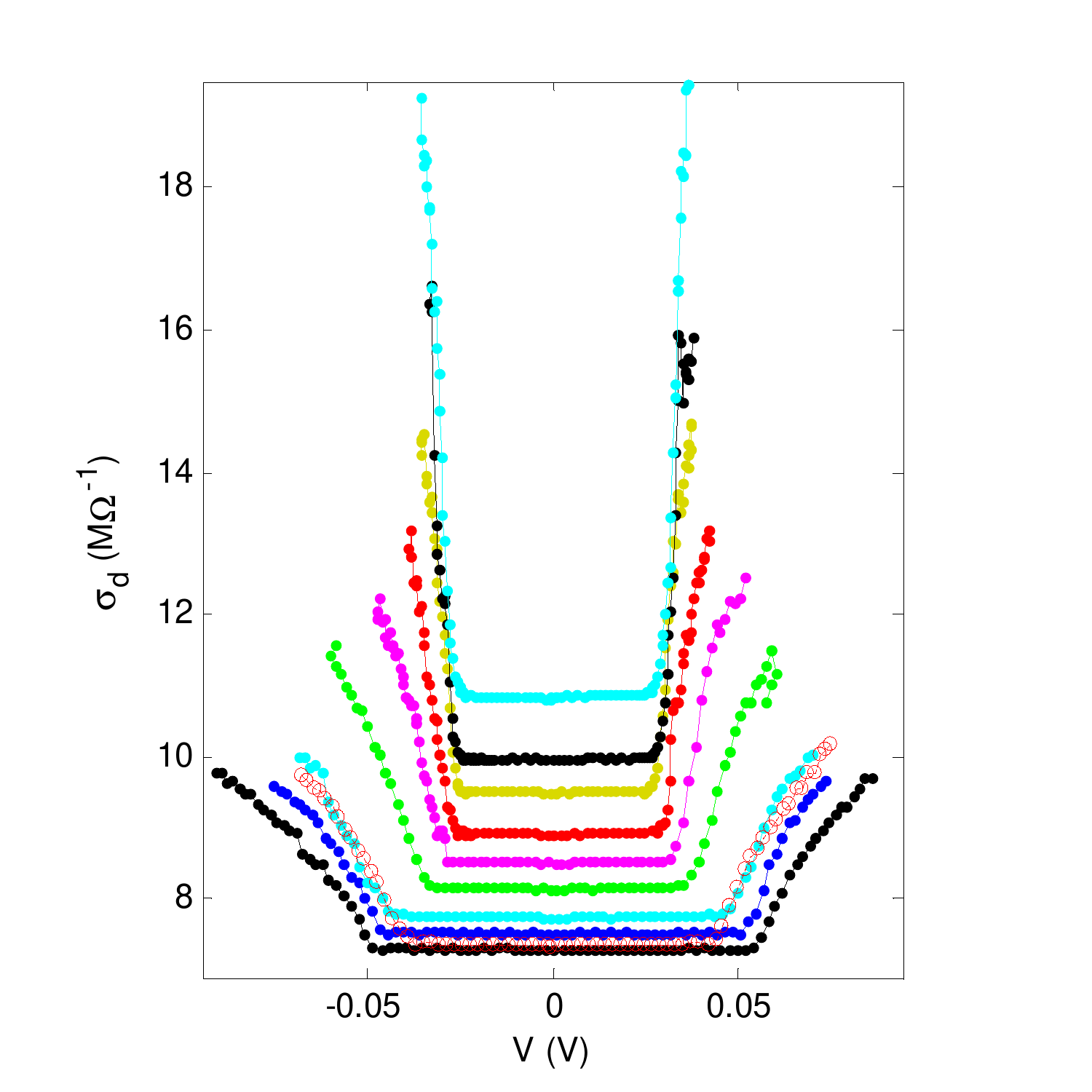}
\caption{The RT voltage dependencies of differential conductivity $\sigma _{d}$ for a sample under uniaxial strain. The strain increases from $\epsilon$ = 0 (the lowest curve) to $\epsilon$ = 1\% (the uppermost curve) in approximately equal steps. The curve marked with red circles was obtained after the strain was removed. The sample length, $L$, is 24$\mu$m.} \label{PD}
\end{figure}

The growth of coherence with strain is also demonstrated by the effect of RF synchronization. At a fixed RF power, the Shapiro steps are more pronounced at increased strain, becoming visible directly in the I-V curves (Fig.\,5). The effect of strain appears similar to that of asynchronous RF irradiation which also improves the coherence of the CDW sliding, as observed for a number of compounds~\cite{3,31,311}, including NbS$_{3}$-II~\cite{3}. However, the origins of enhanced coherence in these two cases are likely different. While RF is believed to periodically draw CDW back to its starting state before it loses coherence~\cite{3}, strain can change the crystal defect structure~\cite{19}. As shown in Fig.\,4, the irreversible growth of coherence after strain removal may be explained by a reduction in the crystal defect density after the uniaxial deformation. For example, it is known that twin-free YBCO crystals can be obtained by applying uniaxial pressure at 420$^{\circ}$C in flowing oxygen~\cite{32}. However, the major part of the coherence growth is reversible (Fig.\,4). This could be attributed to an alignment of the metallic chains under strain. Increased velocities of the internal acoustic modes and their reduced friction in the strained samples can also stimulate coherence of the CDW sliding through CDW-lattice coupling~\cite{8,33}.

\begin{figure}[tbp]
\includegraphics[width=21pc,clip]{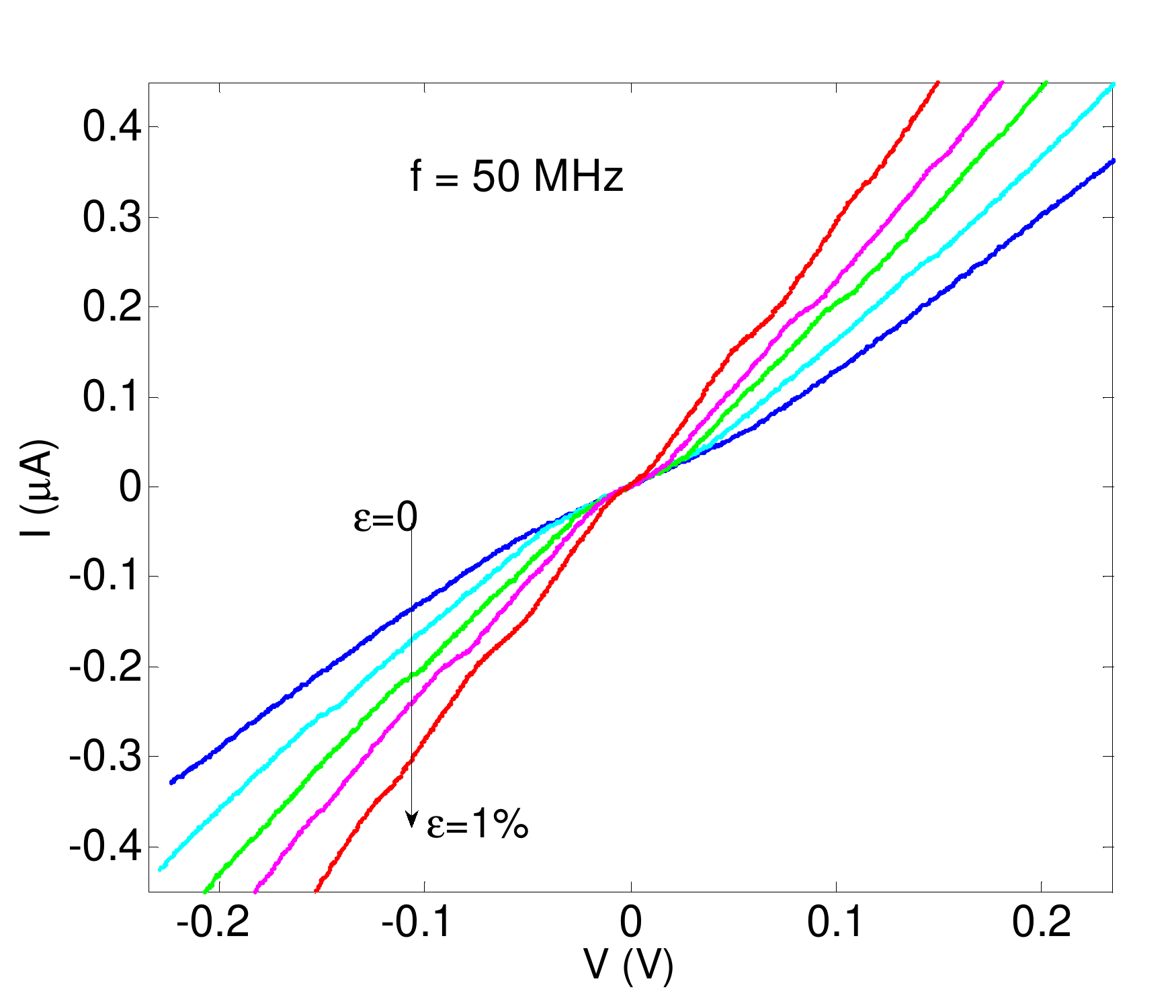}
\caption{RT I-V curves at different strains, $\epsilon$ , under fixed RF irradiation (50 MHz). The sample dimensions are 50$\mu$m $\times$ 0.011$\mu$m$^{2}$. The Shapiro steps become visible with increased $\epsilon$.} \label{PD}
\end{figure}

Studying strain effects of CDW transitions is important to better understand how the CDW condensate forms. Uniaxial strain decreases the anisotropy; as interatomic distances along the conducting chains grow, the interchain distances decrease due to the Poisson contraction. The reduction in anisotropy corrugates the Fermi surfaces and decreases $T_{P}$. At the same time 1D fluctuations are suppressed, which increases $T_{P}$. The actual changes in $T_{P}$ result from the competition between these two effects. Applied uniaxial strain reduces $T_{P}$ in orthorhombic TaS$_{3}$~\cite{5,6} (but with a tendency to increase it after exceeding a critical value of~\cite{30}), NbSe$_{3}$~\cite{5} and K$_{0.3}$MoO$_{3}$~\cite{7}. In the monoclinic TaS$_{3}$ the lower transition is shifted downwards, while no shift was observed for the upper transition temperature up to 1.5\%~\cite{34}.

The effects of uniaxial strain on CDW compounds have been examined less thoroughly than those of hydrostatic pressure, which is in part similar to the effect of stretching. In fact, a general feature of quasi 1D CDW compounds is also a lowering of $T_{P}$ under pressure, indicating that the corrugation of Fermi surfaces has a dominant influence on $T_{P}$~\cite{2,35}. The only known exception is the monoclinic TaS$_{3}$, whose upper transition temperature increases at small pressures. It was proposed that the suppression of 1D fluctuations dominates the $T_{P}$ variation in this strongly anisotropic TaS$_{3}$ polytype.

In the case of NbS$_{3}$-II, strain strongly affects the RT-CDW: $\epsilon$ $\sim$ 1.5\% results in lowering $T_{P1}$ from 360\,K to below 280 K (Fig.\,6). It is also obvious that strain sharpens the CDW transition. The decrease of $T_{P1}$ in NbS$_{3}$-II means that, in spite of the large anisotropy~\cite{36}, it is primarily dominated by Fermi-surface nesting conditions and less by the suppression of the 1D-fluctuations.

\begin{figure}[tbp]
\includegraphics[width=21pc,clip]{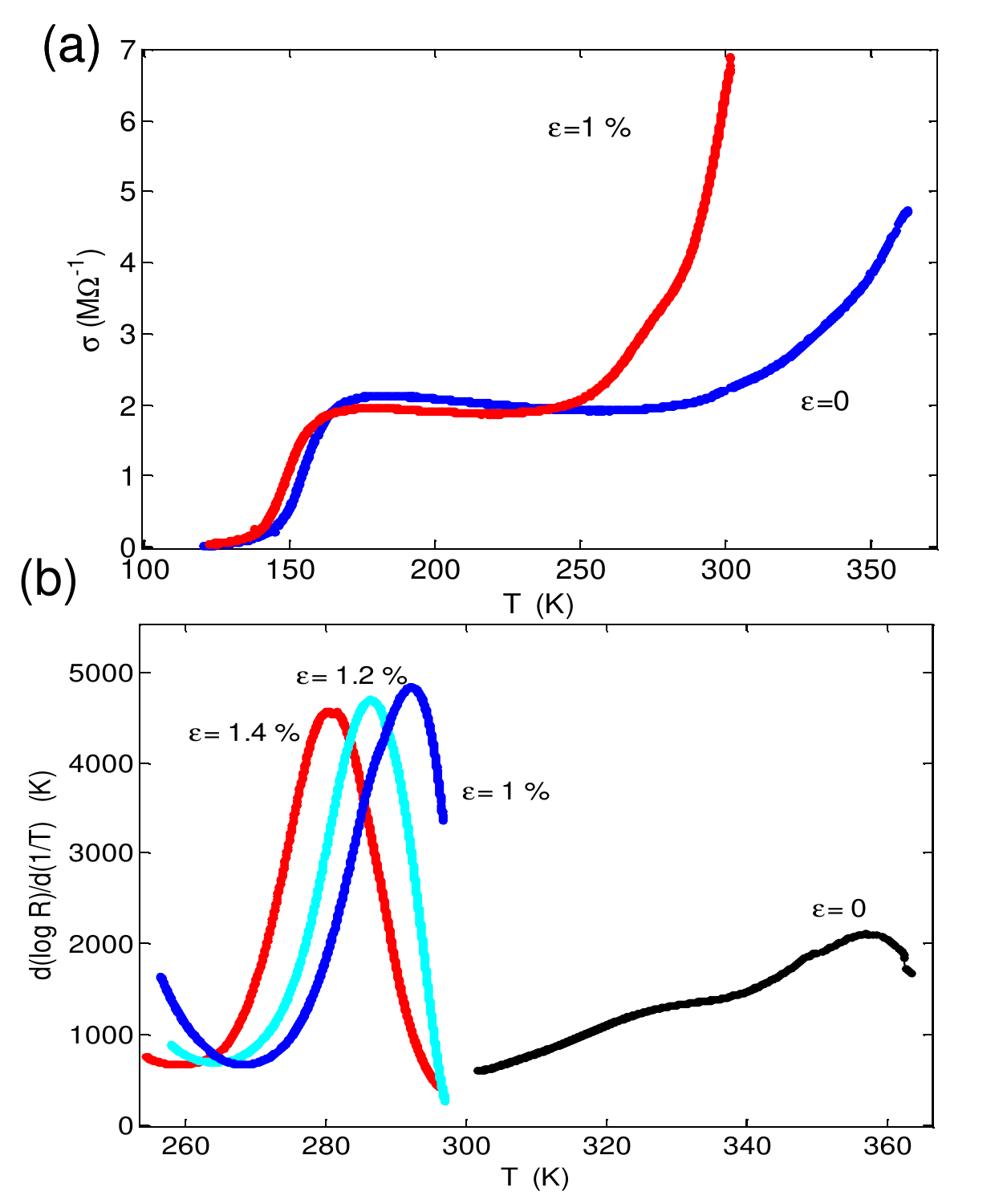}
\caption{ a) The initial $\sigma (T)$ curve (blue) and under ~1\% of stretching strain (red). Both $T_{P1}$ and $T_{P2}$ are lowered under strain. b) The logarithmic derivatives of resistance vs. $T$ for the same sample in the vicinity of $T_{P1}$. The sample dimensions are 22$\mu$m $\times$ 0.014$\mu$m$^{2}$. Measurements under strain were not performed above 300 K due to problems with the epoxy substrates at elevated temperatures.} \label{PD}
\end{figure}

The reduction of $T_{P1}$ with   must be taken into account in the I-V curves under strain (Fig.\,4). For larger applied strains the  $\sigma _{d} (V)$ dependencies were measured close to $T_{P1}$. However, the evolution of the curves with   cannot be attributed to the proximity of $T_{P1}$ only. In the absence of strain, the I-V curves are smeared at temperatures close to the CDW transition~\cite{12}. Therefore, the growth of CDW coherence observed in Fig.\,4, as well as the increased sharpness of the Peierls transition at $T_{P1}$ (Fig.\,6b), are a direct consequence of a tensile strain.

The "inverse" electro-mechanical effects, i.e., the impact of electrical field on the dimensions and the form of the samples, have been less studied for NbS$_{3}$-II. Similarly to orthorhombic TaS$_{3}$, K$_{0.3}$MoO$_{3}$ and (TaSe$_{4}$)$_{2}$I, NbS$_{3}$-II also shows electric-field-induced torsional strain. It can be observed at RT. However, the torsional angles are relatively small and are 1-2 orders of magnitude smaller than those observed in TaS$_{3}$. The most reliable measurement of the torsional angle as a function of voltage is obtained with an ac voltage applied at the mechanical resonant frequency of NbS$_{3}$-II samples. Figure 7 shows two dependences of the torsional angle on current (with and without applied RF irradiation applied) measured with a lock-in technique at the lowest resonant frequency of 3.75 kHz. The applied ac voltage is symmetric, with a gradually sweeping amplitude. The torsional angle was measured optically~\cite{8}. Without RF irradiation, a threshold current for the onset of torsion was observed, like in the cases of the orthorhombic TaS$_{3}$, K$_{0.3}$MoO$_{3}$ and (TaSe$_{4}$)$_{2}$I~\cite{8,37}. RF irradiation (45\,MHz) suppresses the thresholds and increases the torsional angles by an order of magnitude. This effect is in line with the coherence stimulation by an asynchronous RF irradiation~\cite{3}.

\begin{figure}[tbp]
\includegraphics[width=21pc,clip]{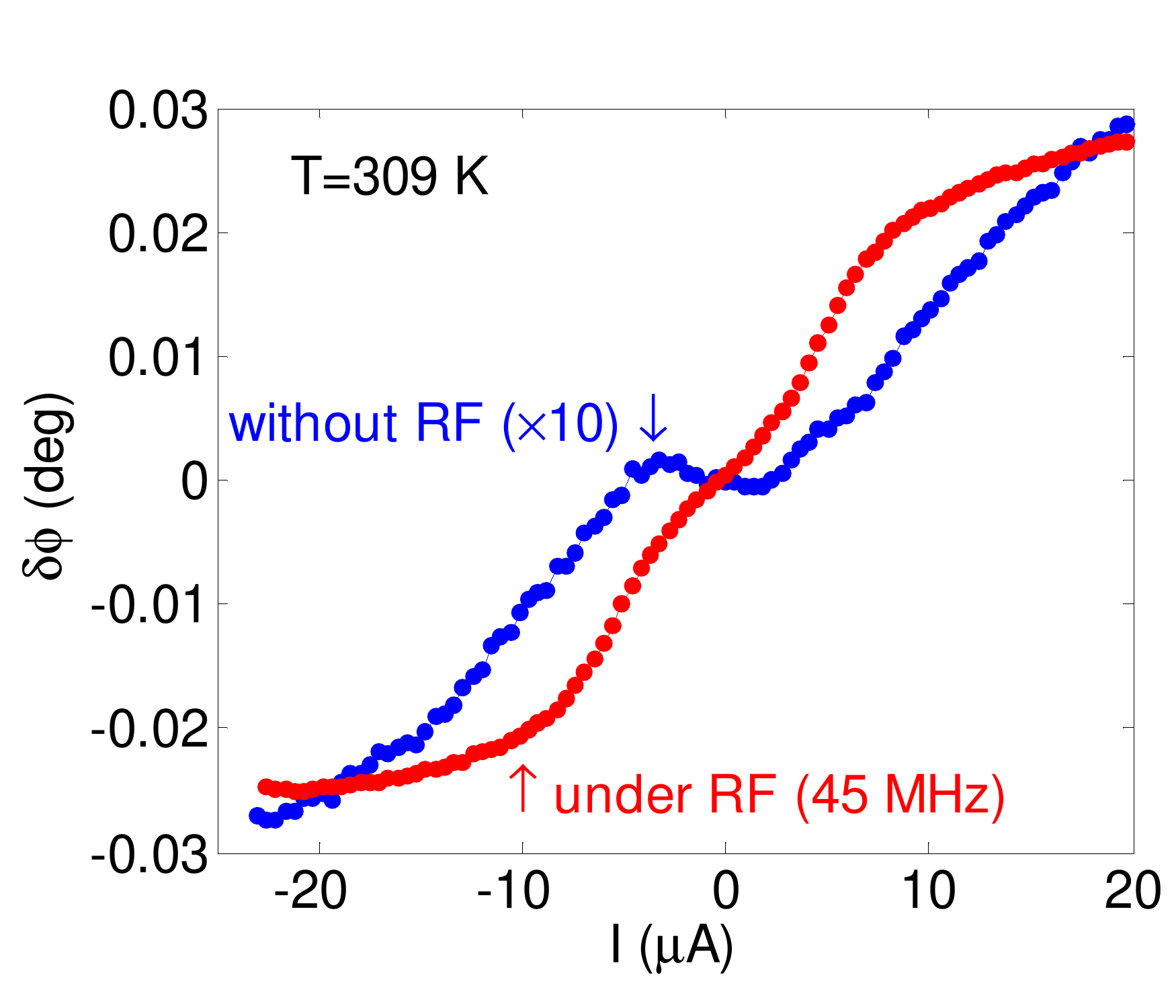}
\caption{ Current dependence of the torsional strain amplitude. The angle is measured with the lock-in technique at a resonance frequency of 3.75 kHz. The current applied to the sample has the form of symmetric meanders. The data without RF irradiation are multiplied by 10. The sample dimensions are 200$\mu$m $\times$ 3.6$\mu$m$^{2}$~\cite{38}} \label{PD}
\end{figure}

\section{THE ULTRA HIGH-$T_{P}$ CDW: INDICATIONS OF A PEIERLS TRANSITION AND A FRÖHLICH MODE.}

\begin{figure}[tbp]
\includegraphics[width=21pc,clip]{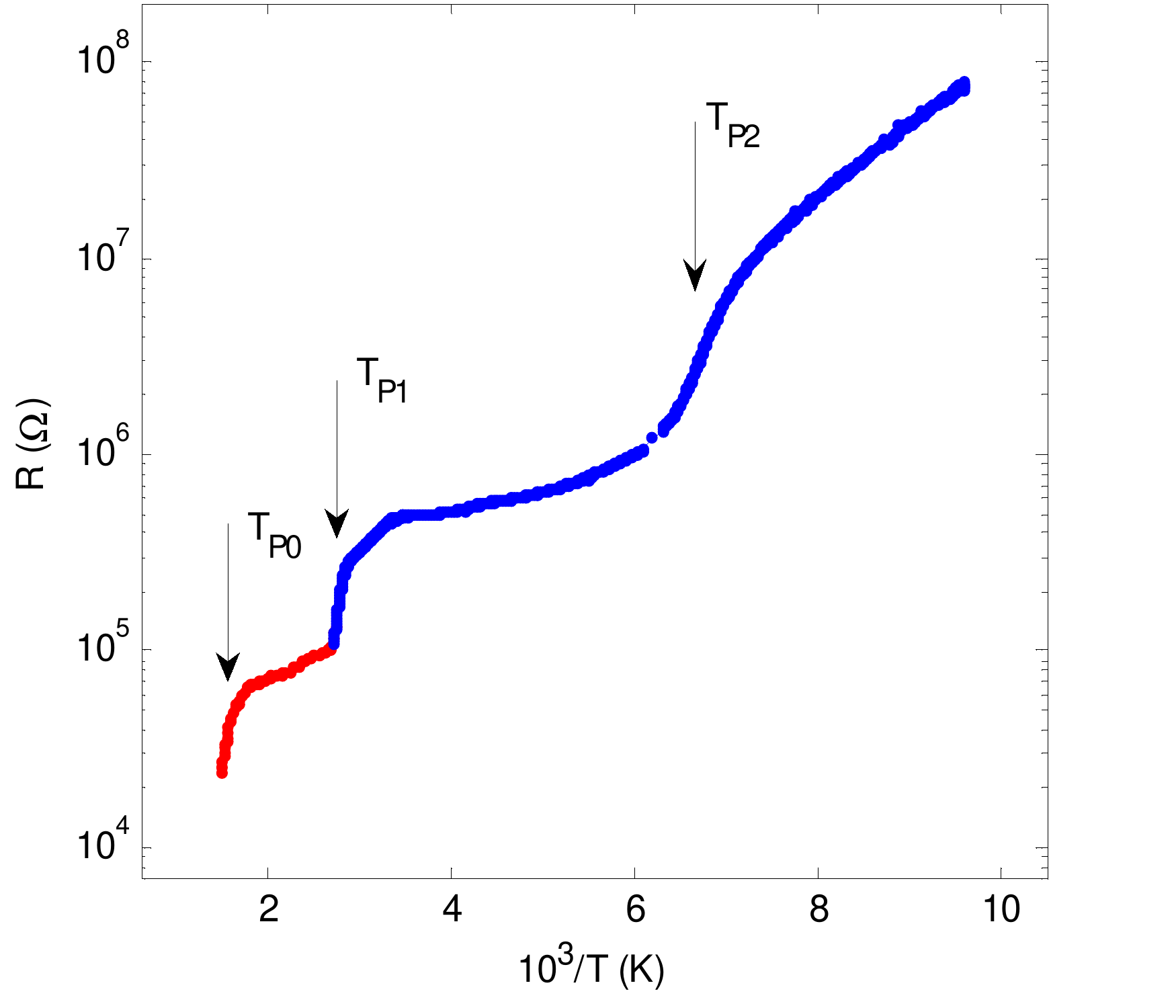}
\caption{ A wide-range temperature dependence of NbS$_{3}$-II resistance. Data from two whiskers (100$\mu$m $\times$ 0.09$\mu$m$^{2}$ - high T, 126$\mu$m $\times$ 0.06$\mu$m$^{2}$ - lower T) are combined into a single graph. The high-temperature points were obtained during a heating cycle in an Ar flow.} \label{PD}
\end{figure}

As mentioned in the introduction section, the diffraction patterns of NbS$_{3}$-II show at 300 K incommensurate satellites belonging to two CDW $q$-vectors: $q_{0}$ = (0.5$a$*, 0.352$b$*, 0) and $q_{1}$ = (0.5$a$*, 0.298$b$*, 0)~\cite{10,11}. While the first set of satellites remain visible up to 450\,K, the second set vanishes gradually above 360\,K, which is close to $T_{P1}$~\cite{11}. When heated above 450\,K under vacuum or ambient conditions, NbS$_{3}$-II crystals start to degrade.

In order to extend the measuring range and prevent crystal degradation, the $R(T)$ measurements at elevated temperatures reported herein were performed in an Ar atmosphere. The temperature was monitored with a thermocouple. Heating to $\sim$ 700\,K and subsequent cooling was performed within several minutes. Figure 8 shows the $R(T)$ curve in a wide temperature range and very clearly reveals the transitions at $T_{P1}$ and $T_{P2}$. The dependence at high temperatures, above $T_{P1}$, was added from a different sample during a heating cycle. Above 600\,K a notable degradation of sample properties begins, resulting in a growth of conductivity. However, during fast cooling from 700\,K the feature in the $\sigma (T)$ reappears in the same temperature range (the dynamic error in temperature determination was tens of K). Therefore the $\sigma (T)$ feature around 620-650\,K is attributed to the transition at $T_{P0}$.

As shown in Fig.\,9, heating to above about 800 K gradually transforms the high-ohmic sub-phase into the low-ohmic one. Further heating transforms NbS$_{3}$-II into a compound with metallic conductivity. As discussed below, the increased conductivity may be attributed to sulfur loss and formation of S vacancies at elevated temperature.

Figure 10 shows a series of $\sigma _{d}(V)$ curves for a NbS$_{3}$-II sample, recorded below and above $T_{P1}$. At $T > T_{P1}$ a gradual growth in conductivity at fields above 0.3 kV/cm is observed (e.g. the curve at 450\,K). It is difficult to separate the effects of CDW sliding and Joule heating in these curves. However, sliding of the UHT-CDW can be checked by the effect of RF irradiation on the $\sigma _{d}(V)$ curves. This effect was studied at RT, where it was possible to place the sample sufficiently close to the RF generator output for better matching. Figure 11 illustrates the $\sigma _{d}(V)$ curves recorded at RT over a wide voltage range. At low voltages, the increase of $\sigma _{d}$ is attributed to RT-CDW sliding, which shows saturation for values above about 1 V. However, at voltages above about 5 V ($ E>$ 10 kV/cm), $\sigma _{d}$ grows rapidly again above the initial saturation level. To check whether this second rise comes from sliding of the UHT-CDW, the effect of coherence stimulation by RF irradiation~\cite{3} was employed. One can see that when the RF field is applied (with all other conditions kept fixed)  $\sigma _{d}$ grows faster and at lower electric fields (Fig.\,11). Thus, the $\sigma _{d}(V)$ diagram exhibits typical features of a CDW conductor with a threshold voltage and a saturation at higher voltages. RF voltage superimposed onto a slowly sweeping DC voltage would give a trivial opposite effect, i.e., smearing out the  $\sigma _{d}$(V) curve (see Fig.\,1 in~\cite{3}). Consequently, the observed increase of $\sigma _{d}$ at higher voltages should be attributed to sliding of the UHT-CDW. The largest current density $j_{c}$ of this sliding CDW, as estimated from Fig.\,11, is $\sim$ 5$\cdot$10$^{6}$\,A/cm$^{2}$; this correspond to a fundamental frequency of $\sim$ 300\,GHz (with $j_{c}$/$f_{f}$=18 A/MHz/cm$^{2}$).

\begin{figure}[tbp]
\includegraphics[width=21pc,clip]{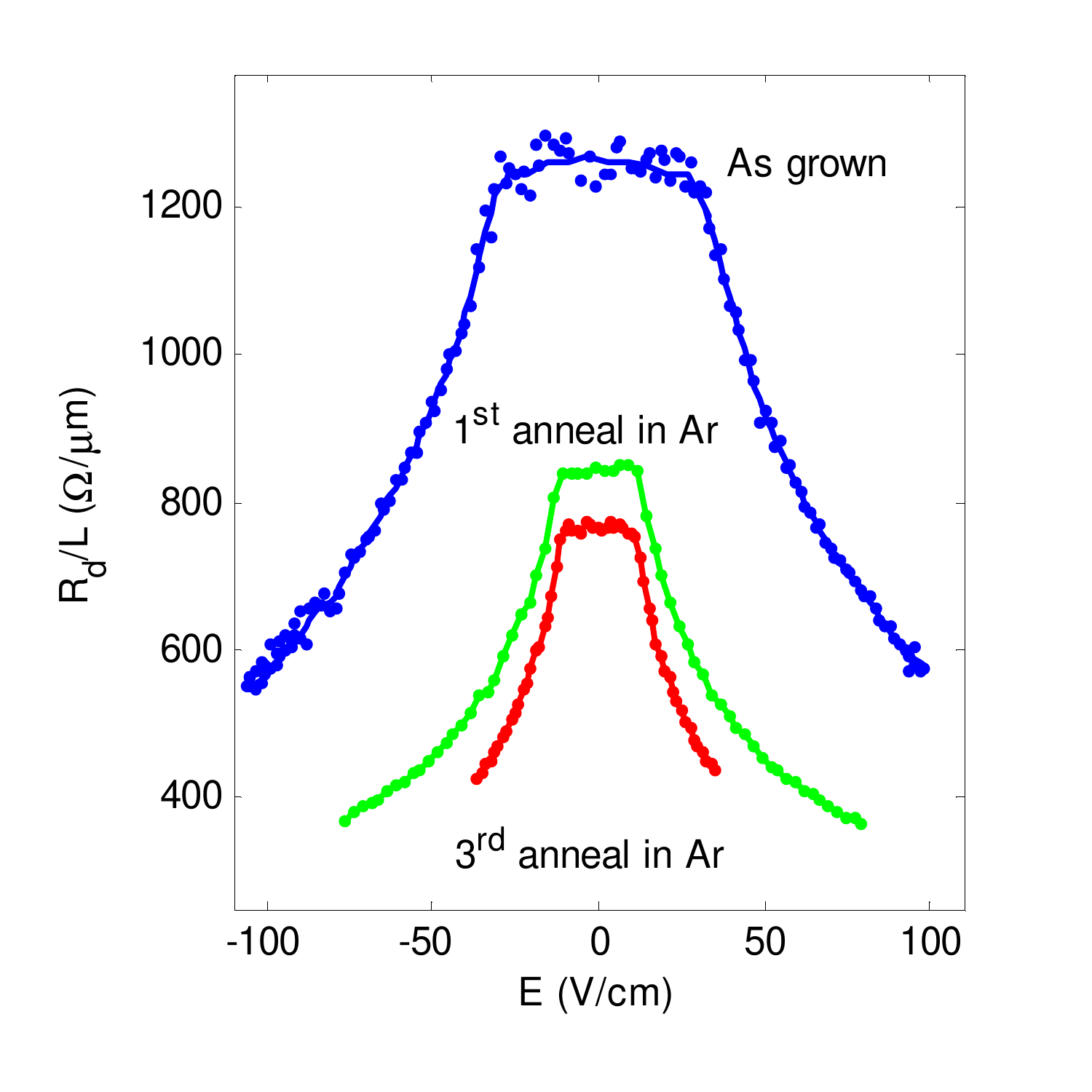}
\caption{ Normalized RT differential resistance vs. electric field of an as-grown high-ohmic sample and after 3 subsequent heating/cooling cycles up to 800-850 K in an Ar atmosphere. The contact-separation L = 1050 $\rightarrow$ 630$\rightarrow$ 252$\mu$m (becoming shorter after each heating cycle because of new contacts).} \label{PD}
\end{figure}

\begin{figure}[tbp]
\includegraphics[width=21pc,clip]{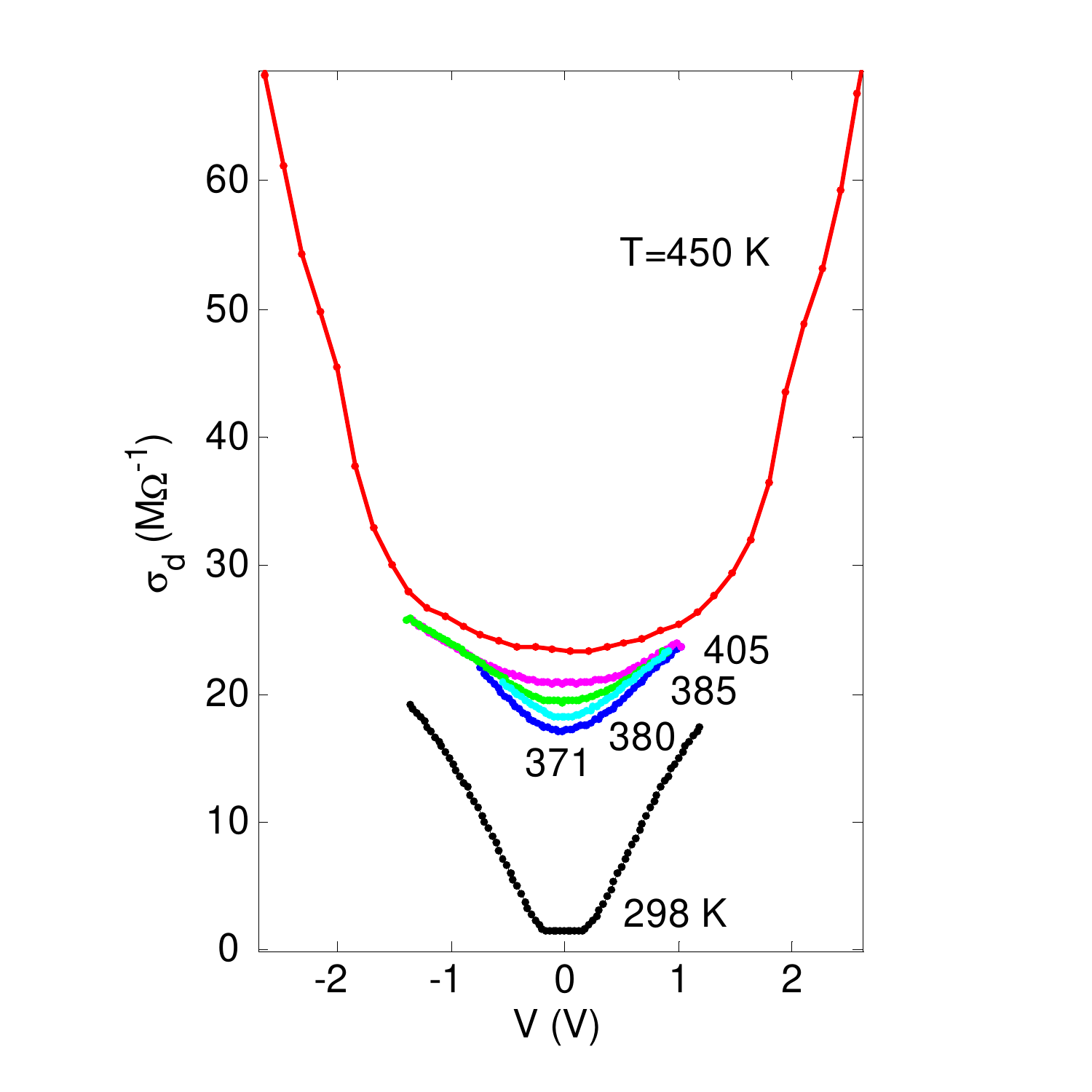}
\caption{ The  $\sigma _{d}$(V) curves for a NbS$_{3}$-II sample below and above $T_{P1}$. The sample length is 44$\mu$m.} \label{PD}
\end{figure}

\section{THE UNUSUAL CDW FORMED BELOW 150\,K.}

\begin{figure}[tbp]
\includegraphics[width=21pc,clip]{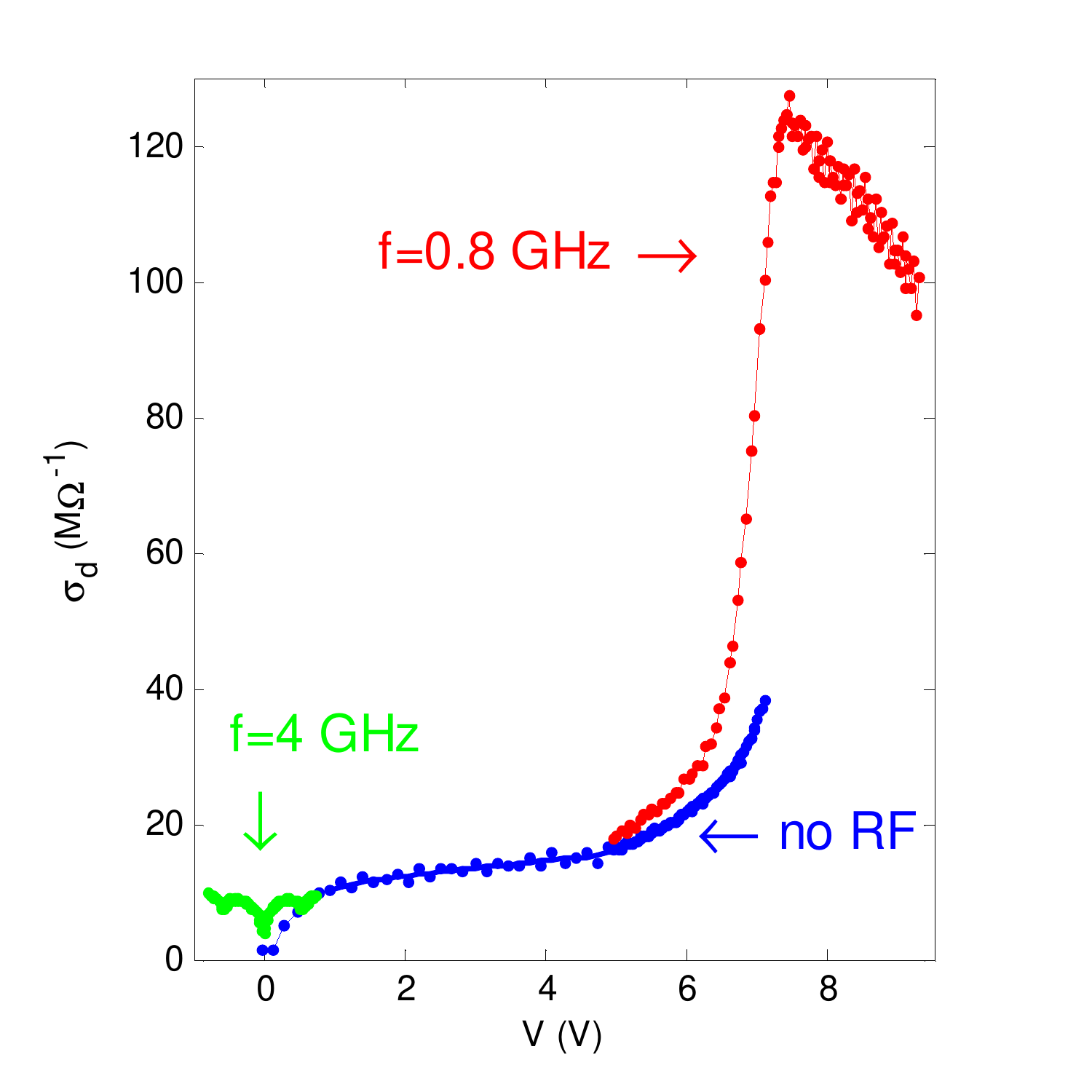}
\caption{ Room-temperature  $\sigma _{d}$(V) curves with and without RF irradiation. Note the Shapiro steps under the 4 GHz irradiation.} \label{PD}
\end{figure}

The transition at $T_{P2}$ = 150\,K (Fig.\,8) remains the least understood. This transition is only observed in the low-ohmic sub-phase, with the drop in specific conductivity, $\sigma _{s}$, near 150\,K being sample dependent~\cite{12,18}. These samples show a threshold in the I-V curves and Shapiro steps above the threshold voltage below $T_{P2}$ (Figs.\,12 and 13), indicating the formation of a new CDW~\cite{12,13,18}. However, the "fundamental ratio" $j_{c}$/$f_{f}$ appears rather low and also sample dependent. The ratio does not exceed 6\,A/MHz/cm$^{2}$ and is 3 times lower than the value for the RT-CDW (18 A/MHz/cm$^{2}$). If the charge density is assumed to be 2$e$ per $\lambda$  on each conducting chain, the highest values of $j_{c}$/$f_{f}$ will correspond to about 1/3 of a chain per unit cell carrying the CDW. This value seems to be close to 1 and one might explain the discrepancy with an inaccuracy of the estimate. However the lowest ratios measured below 150\,K were two orders of magnitude smaller~\cite{12,18}. The difference between the RT-CDW and LT-CDW transport is clearly illustrated by Fig\,12a, where  $\sigma _{d}$ vs. non-linear current is plotted for both CDWs under the same irradiation frequency, 400\,MHz. For the 1$^{st}$ Shapiro step the current of the LT-CDW is nearly 3 orders of magnitude lower than that of the RT-CDW. The LT-CDW thus has an unreasonably low CDW current density for a classical CDW, i.e., for a CDW formed through a Peierls transition.

\begin{figure}[tbp]
\includegraphics[width=21pc,clip]{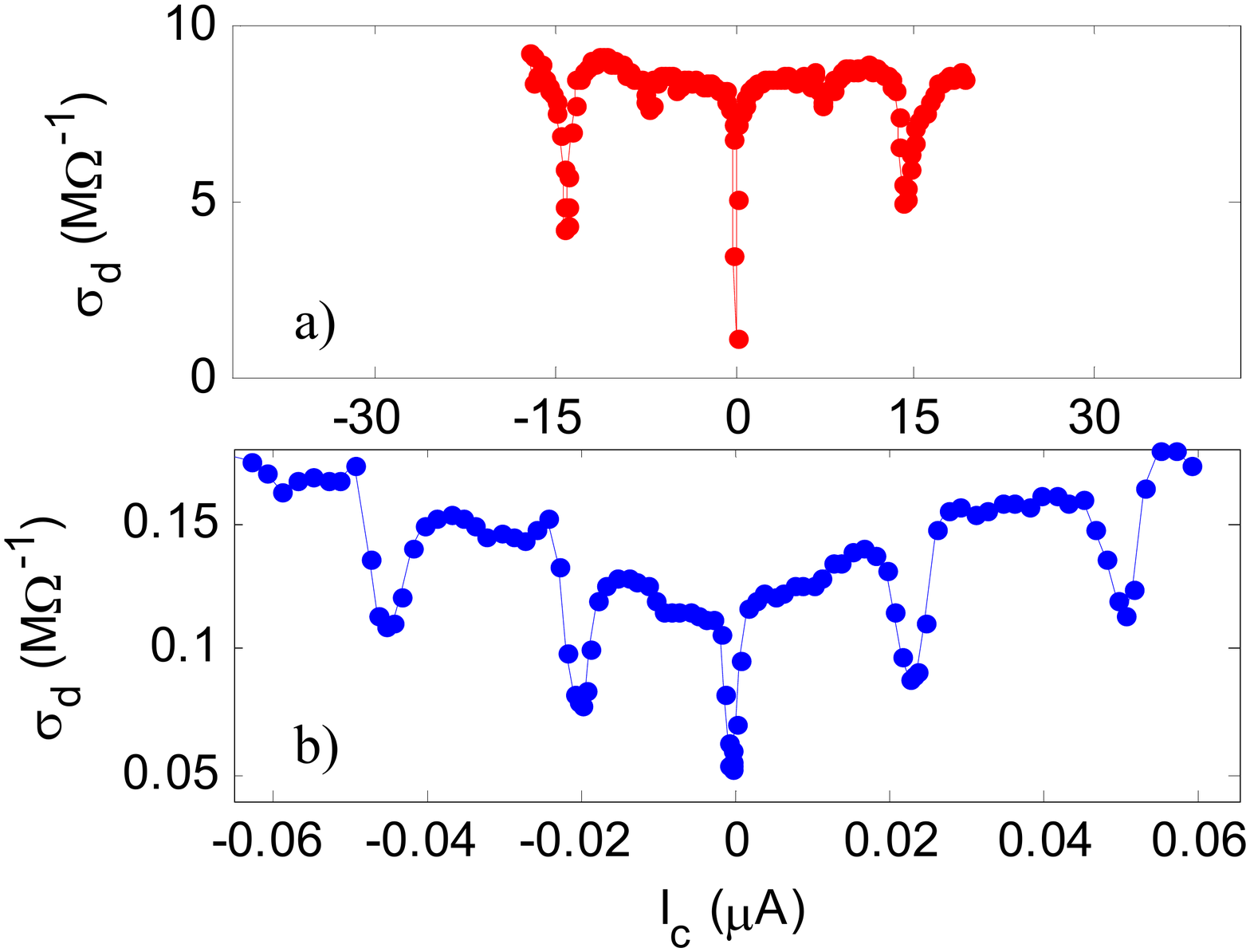}
\caption{ The  $\sigma _{d}$ vs. $I_{c}$ curves under 400 MHz radiation at RT (a) and 122 K (b). The horizontal scales are adjusted to match the positions of 1$^{st}$ Shapiro steps. The dimensions of the sample are 170$\mu$m $\times$ 0.02$\mu$m$^{2}$.} \label{PD}
\end{figure}

Figure 14 shows an Arrhenius plot of specific conductivities, $\sigma _{s}$, for a number of NbS$_{3}$-II samples. For these samples, Shapiro steps at RT allow precise determination of their cross-sectional areas and consequently of their specific conductivities. For comparison reasons the dependence for a NbS$_{3}$-I sample~\cite{16} is added. There is a large variation between the specific conductivities of different samples. The upper group of curves corresponds to the low-ohmic sub-phase and clearly reveals the 150 K transition. This leads to a somewhat wider range of RT specific conductivities for the low-ohmic samples than earlier reported~\cite{8,18}, ranging from 10 to 3$\times$10$^{2}$ ($\Omega$cm)$^{-1}$. Despite the large differences in the actual conductivities, which vary by over an order of magnitude, the temperature dependence $\sigma _{s} (T)$ for the majority of the low-ohmic samples appears very similar on the logarithmic scale. Correspondingly, the drops in $\sigma _{s}$ at $T_{P2}$ appear approximately proportional to their values above $T_{P2}$. Thus, it can be concluded that all the excess electrons, not gapped at $T_{P0}$ and $T_{P1}$, are dielectrized at $T_{P2}$, regardless of the actual electron concentration.

The conclusion is supported by the result presented in Fig.\,15, where the fundamental ratio $j_{c}$/$f_{f}$ of the LT-CDW is shown as a function of the drop in specific conductivity, $\Delta\sigma _{s}$, at $T_{P2}$~\cite{39}. The CDW current density at fixed $f_{f}$ thus appears approximately proportional to $\Delta\sigma _{s}$.

\begin{figure}[tbp]
\includegraphics[width=21pc,clip]{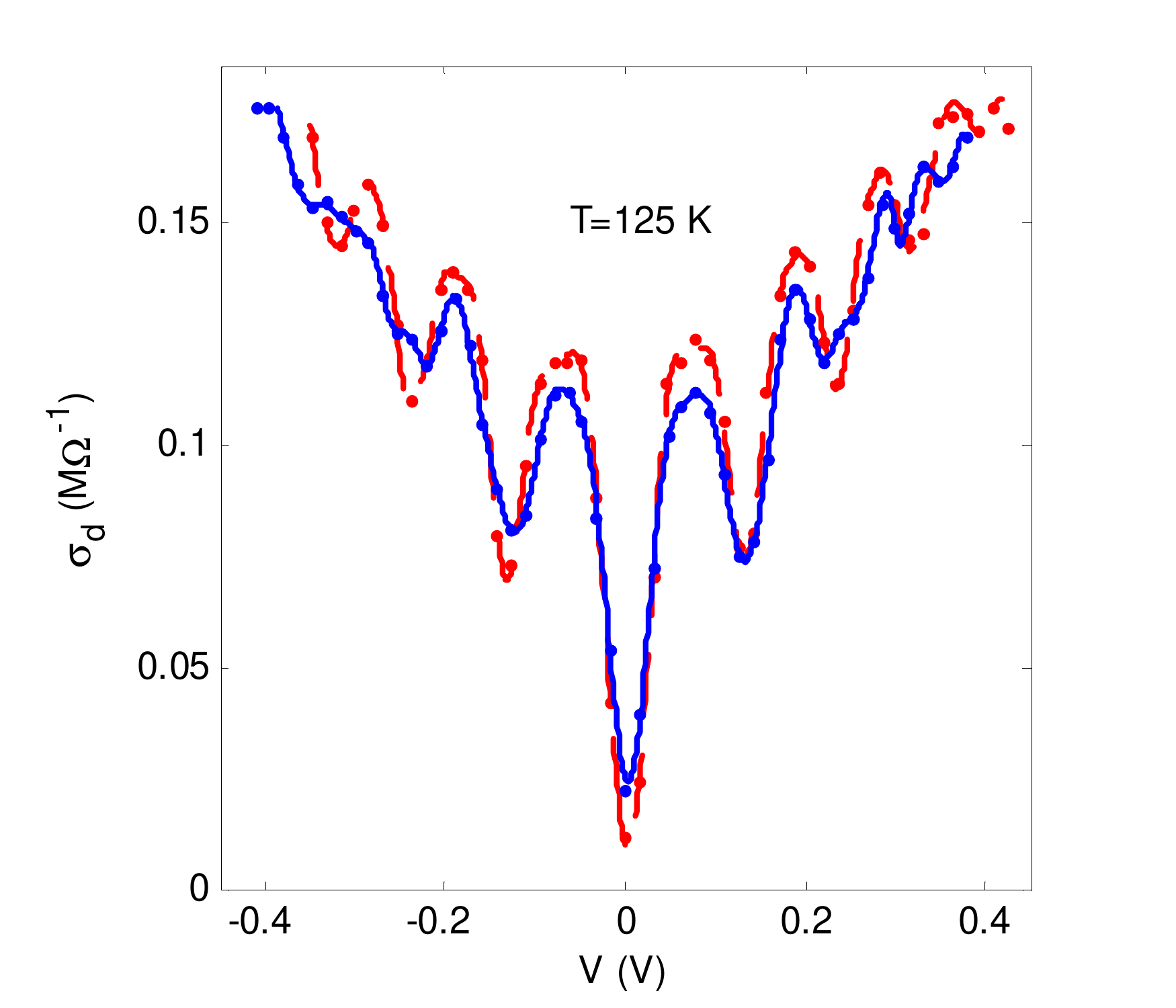}
\caption{$\sigma _{d}$ vs.$V$ curves under 400 MHz radiation below 150 K. The blue (solid) line corresponds to an undeformed sample while the red (broken) curve was obtained from a stretched sample with $\epsilon >$ 1. The dimensions of the sample are 500$\mu$m $\times$ 0.004$\mu$m$^{2}$. The lines are polynomial fits of the experimental points.} \label{PD}
\end{figure}

\begin{figure}[tbp]
\includegraphics[width=21pc,clip]{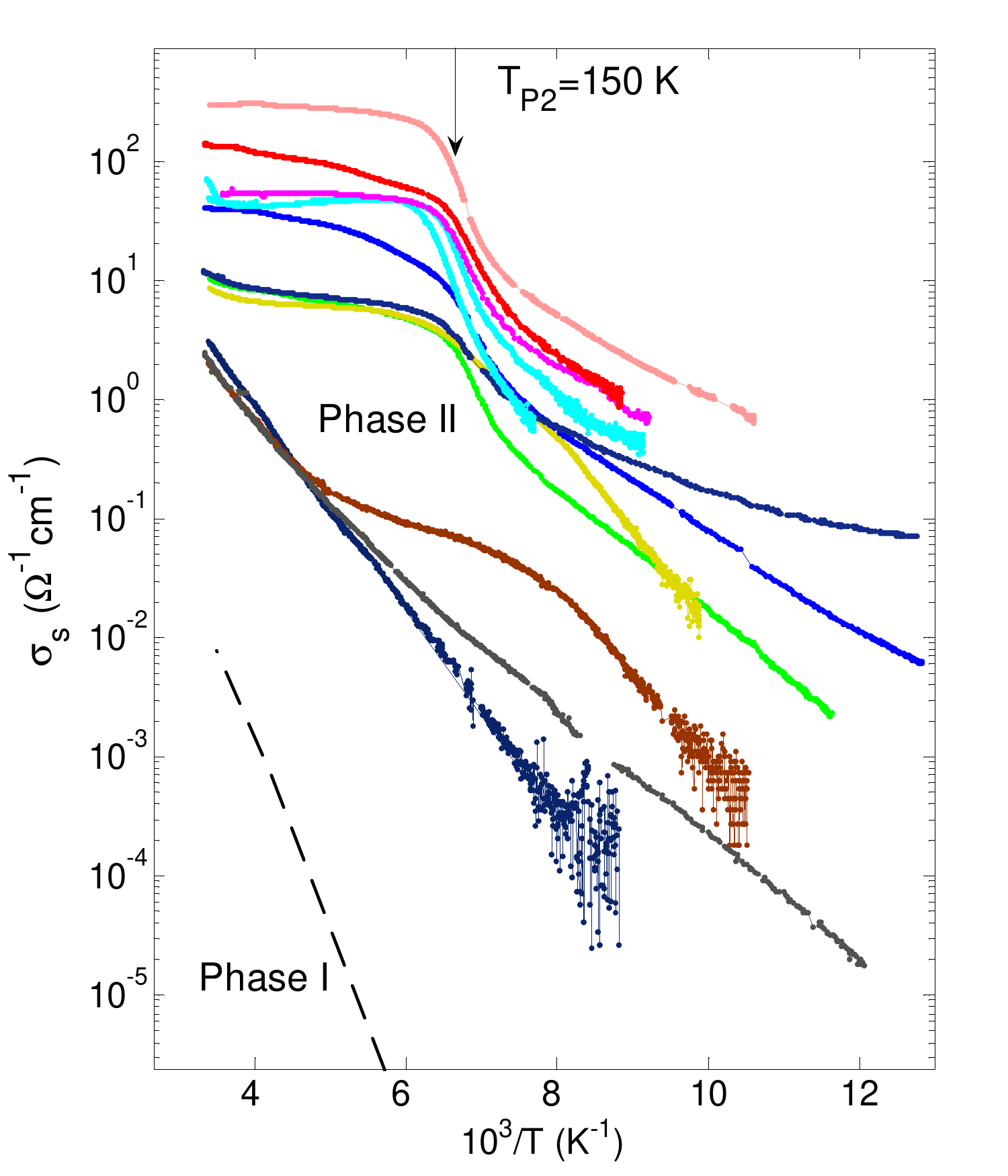}
\caption{ A set of Arrhenius plots of  $\sigma _{s}$ for a number of NbS$_{3}$-II samples. For comparison, a typical  $\sigma _{s}$ (1/T) curve for NbS$_{3}$-I is also shown~\cite{16}.} \label{PD}
\end{figure}

\begin{figure}[tbp]
\includegraphics[width=21pc,clip]{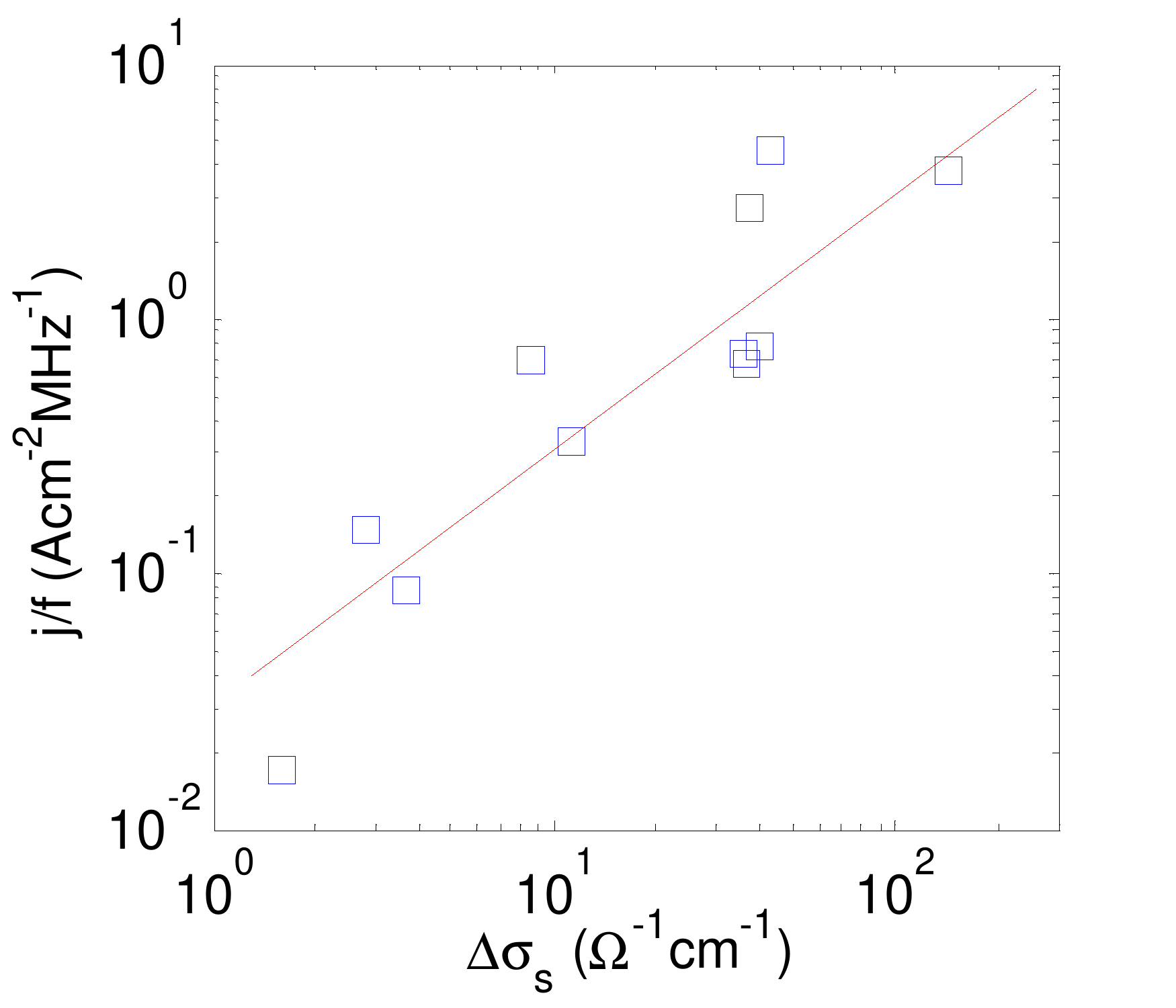}
\caption{ The "fundamental ratio", $j_{c}$/$f_{f}$, of the LT-CDW vs. the specific conductivity drop at $T_{P2}$. The straight line represents a linear approximation of the data.} \label{PD}
\end{figure}

If the single-particle conductivity above $T_{P2}$ and the CDW conductivity below $T_{P2}$ are provided by the same electrons, the mobility in the normal state can be estimated from the relationship between $\sigma _{s}$ and the "fundamental ratio" (Fig.\,15). By multiplying  $\Delta\sigma _{s}$/($j_{c}$/$f_{f}$) with an estimated value of $\lambda$ =10 {\AA}, which corresponds to the wavelengths of the RT- and UHT-CDWs, an electron mobility of about 3 cm$^{2}$/Vs is obtained. For a specific conductivity  $\Delta\sigma _{s}$ = 2.5$\times$10$^{2}$ ($\Omega$cm)$^{-1}$ (see Fig.\,14), this mobility above $T_{P2}$ corresponds to an electron concentration of about 5$\times$10$^{20}$ cm$^{-3}$, which is about 0.3 electrons per unit cell~\cite{1,2}.

Samples with $\sigma _{s}$(300\,K) $<$ 10 ($\Omega$cm)$^{-1}$ belong to the high-ohmic sub-phase. Their pronounced dielectric behavior of the  $\sigma _{s}(T)$ curves at RT and below (Fig.\,14) indicates that the free carriers arise from thermal excitations across the Peierls dielectric gap formed at $T_{P1}$. A pure high-ohmic sample shows an activation energy of about 2000\,K below $T_{P1}$. This value is close to 2500\,K, the half-value of optical gap recently reported for NbS$_{3}$-II~\cite{40}. Evidently, the number of free electrons in these samples is insufficient to condensate into a collective state. Alternatively, the transition can just become invisible because of the insufficient electron concentration.

\begin{figure}[tbp]
\includegraphics[width=21pc,clip]{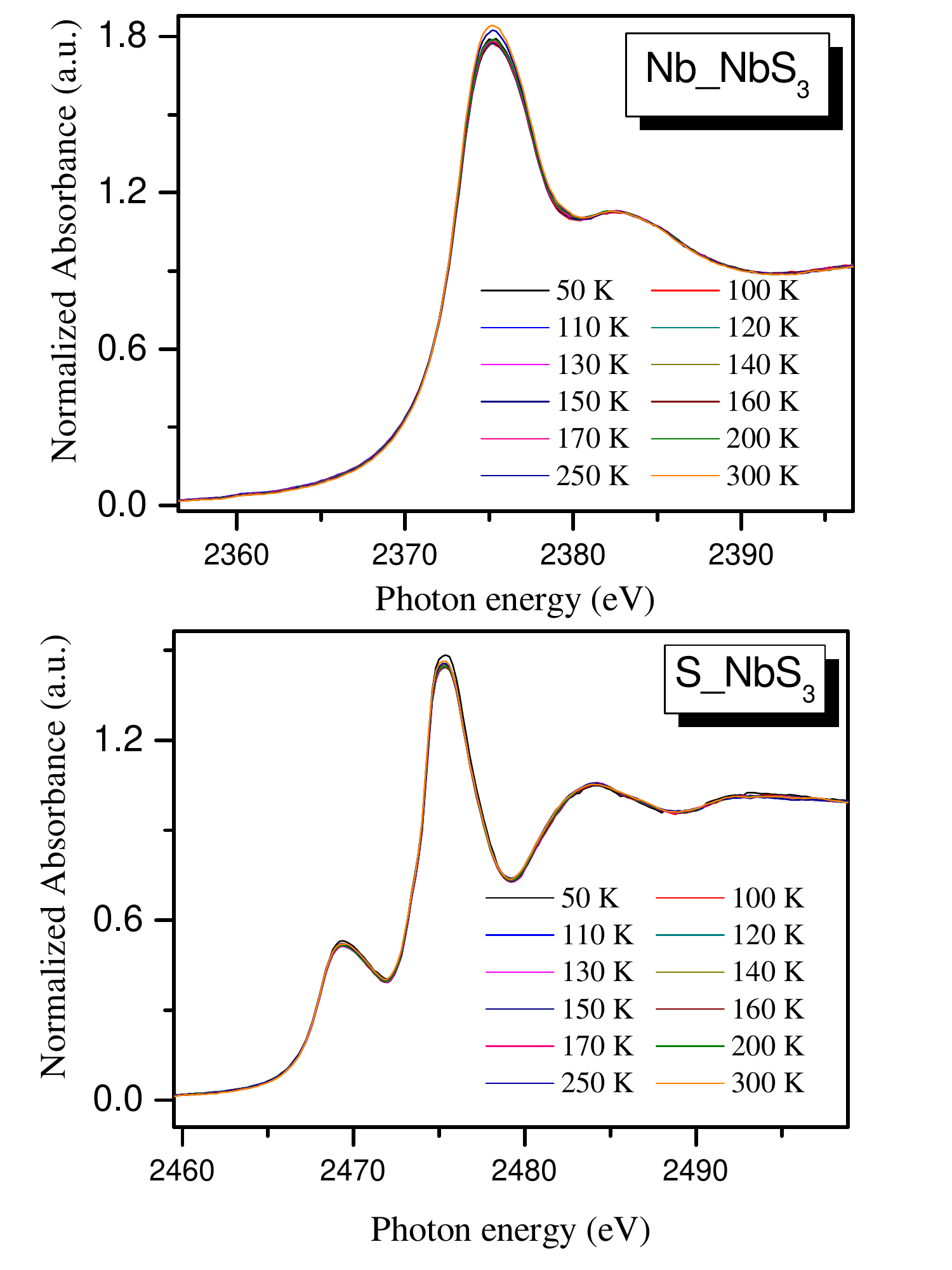}
\caption{ Fluorescene-detected XANES Nb L3-edge (a) and S K-edge (b) spectra between 300 K and 50 K} \label{PD}
\end{figure}

To better understand the properties of NbS$_{3}$-II, we determined the chemical composition of these samples with electron-probe microanalysis (EPMA). Reliable compositions were only obtained for samples with transverse dimensions larger than 1 $\mu$m, which all belonged to the low-ohmic sub-phase. It was thus impossible to establish a difference in composition between the two sub-phases. From a total of 15 measurements at several sample locations, we obtained a S:Nb ratio of 2.87$\pm$0.04 for the low-ohmic sub-phase. Two independent studies, performed at NTU, Taiwan and MPTI, Russia, gave comparable results.

The significant non-stoichiometry in NbS$_{3}$-II indicates either the presence of S vacancies or an excess Nb in the low-ohmic samples. It has been reported previously that NbS$_{3}$ is susceptible to S loss with heating~\cite{41}. It is therefore likely that the variation in specific conductivity found in the NbS$_{3}$-II samples is a result of S vacancies acting as donors, similar to the observed behavior of TiS$_{3}$~\cite{42,43} and TiSe$_{2}$~\cite{44,45}. Consequently, the higher concentration of S vacancies would account for the reduced electrical resistivity observed in samples, belonging to the low-ohmic sub-phase. Although the EPMA measurements were not performed on the high-ohmic samples, it is reasonable to assume that these samples are closer to the stoichiometric composition. Their transformation into the low-ohmic phase under high-temperature treatment (Fig.\,9) is then connected with a loss of S.

\begin{figure}[tbp]
\includegraphics[width=21pc,clip]{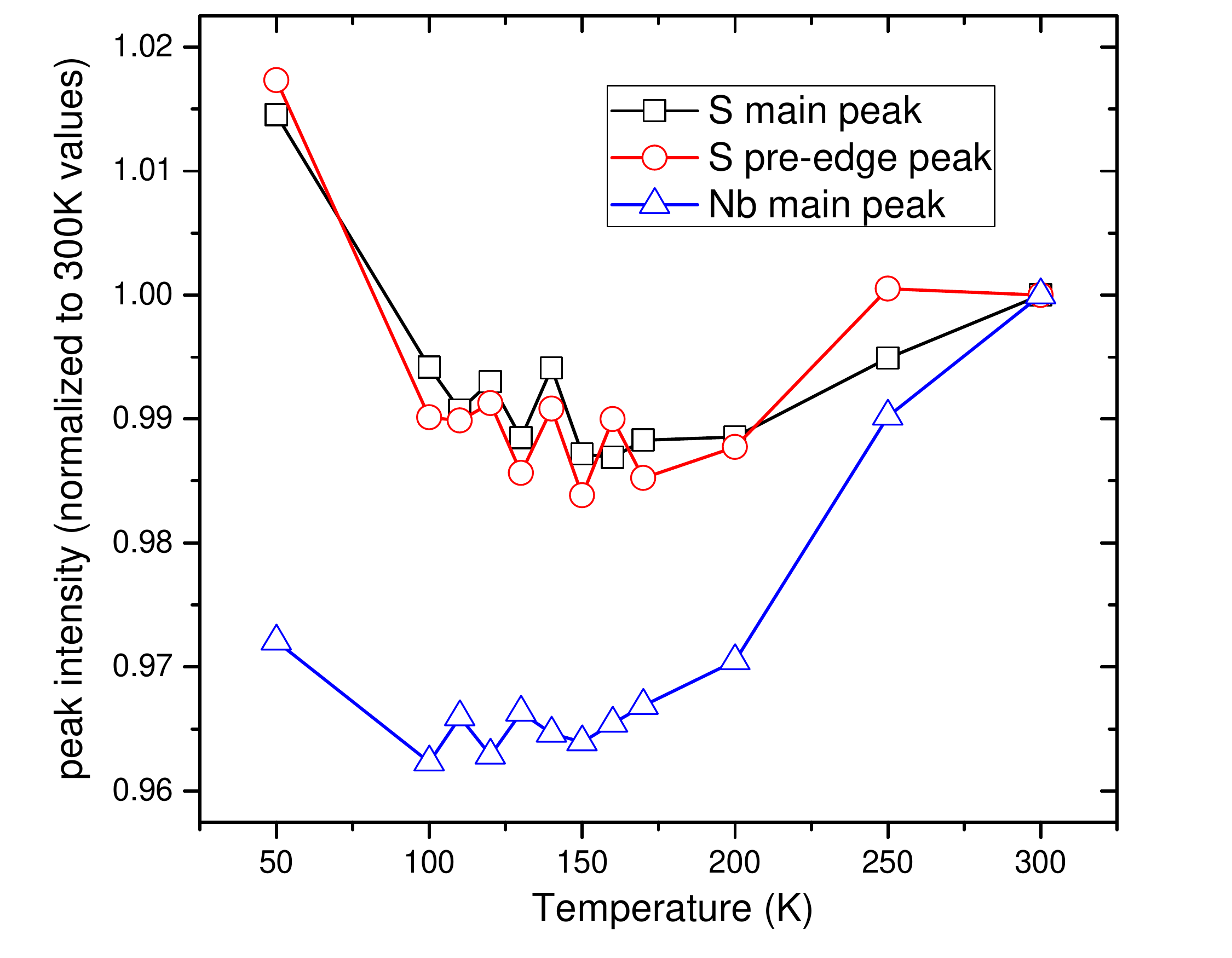}
\caption{ The normalized intensities of the S K-edge, pre-edge and Nb L3-edge as function of temperature.} \label{PD}
\end{figure}

The LT-CDW seems to further condense electrons from a state already dielectrized by the RT-CDW. This is rather unusual. For a conventional RT-CDW, condensation of the Nb $d$-state electrons leads to a Peierls gap. To further condense electrons in the Nb $d$-state at $T_{P2}$, some additional electrons can be transferred from the S $p$-state into the Nb $d$-state by forming thus $(S_{2})^{2}-$ pairs. We have used X-ray near edge absorption spectroscopy (XANES) to probe the hole occupation in Nb $d$-state and S $p$-state from fluorescence-detected Nb L3-edge (at 2375\,eV) and S K-edge(at 2476\,eV) absorption lines. From RT to 150\,K, the Nb-L3 peak intensity decreases due to reduced hole occupation of the Nb 4$d$ state, as shown in Fig.\,16(a). Similarly, the pre-edge feature (2470\,eV) of the S K-edge also decreases in intensity between these temperatures, as shown in Fig.\,16(b). This pre-edge arises from the transition from 1s to a $p$-$d$ mixed empty bound state when the S 3$p$ state takes on a "hole" character by mixing with the Nb 4$d$ states. A reduction of the S pre-edge intensity is consistent with the expectation that the Nb 4$d$ hole occupation is reduced. However, we also found that the S K-edge main peak decreases in intensity by lowering the temperature from 300\,K to 150\,K. This means that a simple scenario of electrons transferred from S 3$p$ to Nb 4$d$ states is questionable. Instead, some electrons are transferred to both the S 3$d$ and Nb 4$d$ states. Fig.\,17 shows the evolutions of the Nb L3-edge peak, the S K-edge peak, and the S K-edge pre-edge intensities as functions of temperature. We note that the behaviour shown in Fig.\,17 is reversible with temperature.

A source of the electrons to occupy Nb 4$4$ and S 3$p$ states could be the S vacancies acting as electron donors. The concentration of S vacancies is much larger than typical doping concentrations in semiconductor materials and therefore likely makes NbS$_{3}$-II a degenerate semiconductor. If the electronic structure of NbS$_{3}$-II is such that an electron pocket (from the Nb 4$d$ state) and a hole pocket (from the S 3$p$ state) both exist at the Fermi energy, a slight shift of it can either decrease or increase the hole occupations of both Nb 4$4$ and S 3$p$ states. Such an electronic structure is depicted in Fig.\,1(a) of Ref.~\cite{47} and is believed to exist in WTe$_{2}$~\cite{46}. As NbS$_{3}$-II is cooled from 300\,K to 150\,K, thermal generation of carriers becomes less important, the Fermi level tends to move upwards, towards the donor impurity band. This is revealed by the decrease of both the Nb L3-edge and S K-edge peak intensities observed in XANES.

By reducing the temperature from 150\,K to 50\,K, we observe that the Nb L3-edge and the S K-edge peak intensities recover, resulting in a broad minimum at $T_{P2}$. While the exact cause of this phenomenon is still uncertain, the removal of electrons from Nb 4$4$ and S 3$p$ states could be related to exciton formation. Reduced screening is expected in low-dimensional materials and at lower temperatures, when free carriers are fewer; this leads to enhanced exciton binding energy. If the LT-CDW has indeed an excitonic insulator nature, exciton formation would remove electrons below $T_{P2}$ from the Nb 4$4$ and S 3$p$ states. This leads to the XANES-observed recovery of Nb L3-edge and S K-edge peaks' intensities below $T_{P2}$.

We also observed a similar minimum of the Ta L3-edge XANES in the orthorhombic phase of TaS$_{3}$ near $T_{P}$ = 220\,K, which is believed to be a "classical" Peierls transition~\cite{1}. However, in o-TaS$_{3}$ the S K-edge was not simultaneously measured with the Ta L3-edge. We therefore do not yet know whether o-TaS$_{3}$ and NbS$_{3}$-II exhibit similar XANES behaviours. This comparative measurement should be undertaken in the future.

\begin{figure}[tbp]
\includegraphics[width=22pc,clip]{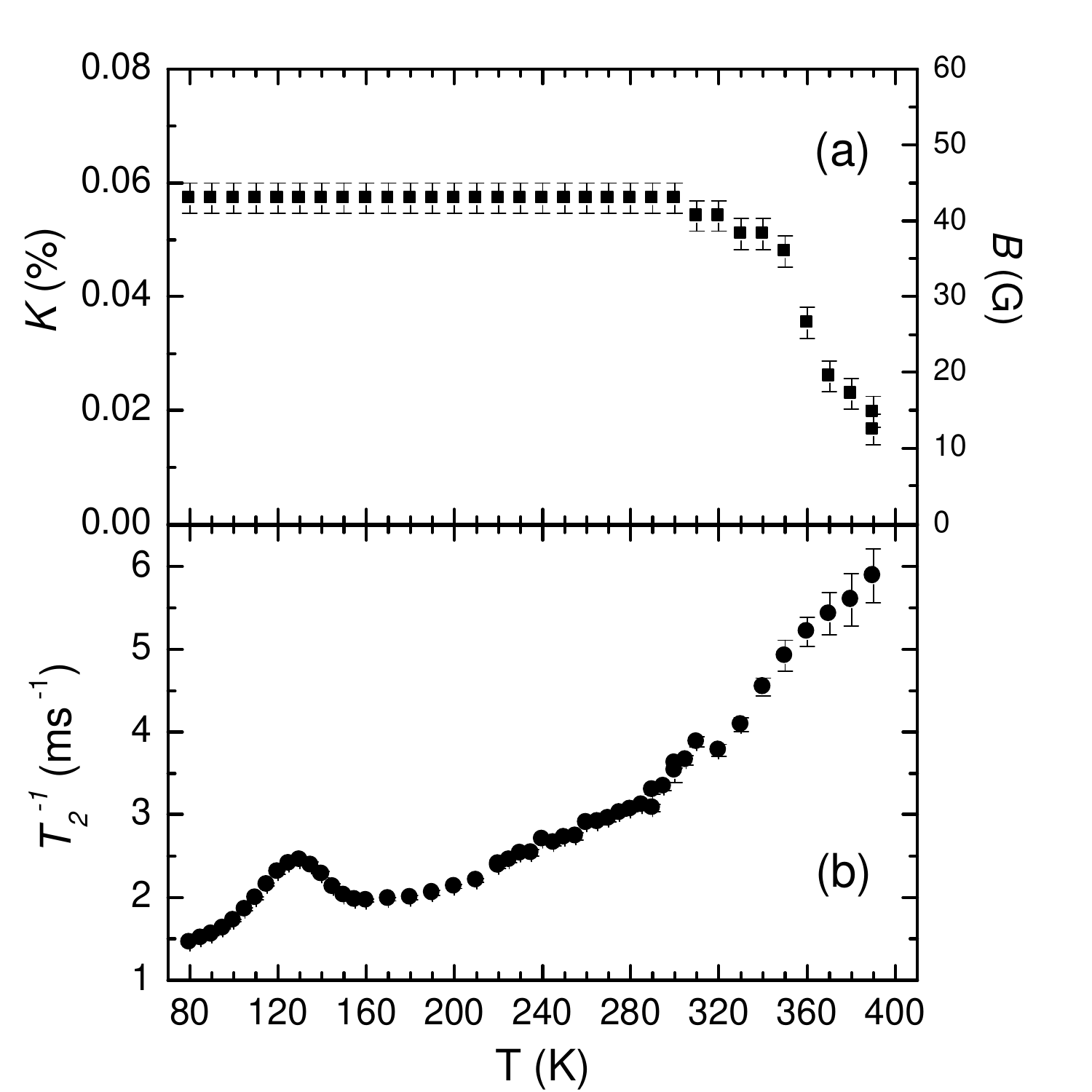}
\caption{ a) Temperature dependence of the Knight shift of the $^{93}$Nb NMR central line. The corresponding "extra" effective magnetic field H at the $^{93}$Nb nuclear sites is shown on the right scale (the external fixed field $B_{0}$=7.5 T). b) Temperature dependence of the nuclear transverse magnetization relaxation rate, $T_{2}^{-1}$.} \label{PD}
\end{figure}

In addition to XANES, preliminary data from STEM-based electron-energy loss spectra (EELS) suggest the local environments of S atoms change above and below the transition. At 105\,K the L-edge of S is characterized by a single peak at 162\,eV, while at 290\,K an additional peak is formed at 167 eV. Meanwhile, the M2 and M3 edges of Nb do not show noticeable changes. The STEM-EELS data support the suggestion that there is a change in the local environment of sulfur atoms. It is possible that changes in the S positions lead to a change in the number of S-S bonds forming S$(_{2})^{2-}$ instead of 2S$^{2-}$ states, or to a partial activation of S donor vacancies at low temperatures. These processes would provide another "degree of freedom" for the S atoms to change the number of free electrons without altering the Nb:S ratio.

NMR studies provide an additional insight into the $T_{P1}$- and $T_{P2}$- transitions. The $^{93}$Nb spectrum of non-oriented NbS$_{3}$-II powder samples in a fixed magnetic field $B_{0}$=7.5\,T clearly shows the presence of many inequivalent Nb sites in this phase. The temperature dependence of the most intense peak of the central line (which corresponds to the -1/2 $\leftrightarrow$ 1/2 transitions) allows clear detection of the CDW transition at $T_{P1}$. In Fig.\,18(a) the Knight shift values, $K$ = ($f$/$f_{0}$-1)$\bullet$100\%, of this central line in the temperature range 80 K-390\,K are shown. Here $f$ is the central line peak frequency, and $f_{0}=\gamma\bullet B_{0}/2\pi$ is the reference frequency, which is proportional to the external magnetic field $B_{0}$ and to the $^{93}$Nb gyromagnetic ratio $\gamma/2\pi$=10.4065\,MHz/T. The Knight shift corresponds to an "extra" effective field at the nuclear site from the polarization of the conduction electrons in the presence of an external field. The right scale in Fig.\,18(a) shows the estimated values of this "extra" field. Therefore, Fig.\,18(a) reveals a new ordered state below $T_{P1}$. The most obvious reason for this is a lattice distortion coupled with CDW formation, as observed in NbSe$_{3}$~\cite{48}.

There is no change in the Knight shift around $T_{P2}$ = 150\,K (Fig.\,18(a)). However, studies of the nuclear relaxation do reveal a feature in this temperature region. The temperature dependence of the nuclear transverse magnetization relaxation rate, $T_{2}$$^{-1}$, measured on the same most intense peak of the central $^{93}$Nb NMR line, is shown in Fig.\,18(b). A maximum of $T_{2}$$^{-1}(T)$ appears at about 130\,K. The loss (decoherence) of transverse nuclear magnetization happens due to different time-dependent local magnetic fields at nuclear sites. The $T_{2}$$^{-1}(T)$ dependence can be explained by the assumption that the characteristic time of such microscopic fluctuations causing relaxation is increasing as temperature decreases. At low temperatures such fluctuations slow down and their correlation time becomes comparable to the spin-echo time (tens of microseconds). In such cases, decoherence of the transverse nuclear spin magnetization becomes faster and the corresponding transverse magnetization relaxation time $T_{2}$ becomes shorter. At even lower temperatures the fluctuations become very slow and their characteristic time appears much longer in comparison with the time of spin-echo formation. Consequently, the fluctuations do not contribute anymore to the transverse magnetization relaxation and $T_{2}$$^{-1}$ begins to decrease again. Thus, the maximum in $T_{2}$$^{-1}$ at 130\,K corresponds to a "freezing" of one of the fluctuation sources causing relaxation~\cite{49,50}. The most probable candidate for such a source is an electronic or a lattice distortion, which exists at $T_{P2}$ or even above it. Such a behavior can be expected in case of strong 1D fluctuations, where the transition signifies 3D ordering of the CDW fluctuations (see Ref.~\cite{51}, e.g.). A similar behavior of $T_{2}$$^{-1}$ was reported for the well-known CDW conductor NbSe$_{3}$ near 130\,K~\cite{49}, i.e. somewhat below the 1$^{st}$ CDW transition, 145\,K.

Unlike the Knight shift, the relaxation of the transverse magnetization can be stimulated by fluctuation of both magnetic and electric fields. Therefore, the maximum in $T_{2}$$^{-1}(T)$ at 130\,K can reveal not only a lattice but also an electronic ordering. Moreover, one can suppose that the LT-CDW is mainly an electronic ordering, with lattice distortion remaining a small, secondary effect. This is consistent with the absence of changes in the Knight shift at $T_{P2}$, as well as the failed attempts to detect a lattice distortion at 150\,K, either with electron~\cite{10} or X-ray powder diffraction~\cite{22}.

As discussed, uniaxial strain can be considered a probe for the Peierls transition. Unlike $T_{P1}$, $T_{P2}$ appears much less sensitive to strain (Fig.\,6a). Some lowering of $T_{P2}$ can only be detected with highest applied strains. Given the same strain, the relative reduction of $T_{P2}$ is an order of magnitude smaller than that of $T_{P1}$. Since the suppression of the fluctuations increases $T_{P2}$ and the corrugation of Fermi surfaces has the opposite influence, a balance of these two effects in LT-CDW might explain why $T_{P2}$ is less sensitive to strain than $T_{P1}$. We also note that different behaviors with respect to strain were observed in other trichalcogenides with multiple CDW transitions. In the monoclinic TaS$_{3}$, the lower transition appears to be more sensitive to strain~\cite{34} and pressure~\cite{2,35}, as compared to the upper one. In NbSe$_{3}$, another compound with two CDWs, the effects of pressure and strain are notably different; while under pressure the lower transition temperature decreases faster than the upper one~\cite{2,35}, the rate of decrease is much slower under uniaxial strain, especially in the range of small strains~\cite{5}. In the case of NbS$_{3}$-II we can suppose that either $T_{P2}$ is more dominated by fluctuations than $T_{P1}$, or the transition at $T_{P2}$ has a different origin.

The same effect of strain is observed for the non-linear conduction of the LT-CDW (Fig.\,13): the I-V curves do not show notable changes up to about 1.5\% strain (the apparent slight improvement of coherence was not repeated in other samples). The weak dependence of the LT-CDW on strain can indicate weak coupling of this CDW with the lattice. In view of this, the low sensitivity of $T_{P2}$ on strain would also be rather coupled with the special nature of the LT-CDW, than to the effect of 1D fluctuations.

\section{DISCUSSION.}

The results presented show that NbS$_{3}$-II appears to be a unique quasi 1D compound. It shows two high-temperature CDW transitions at $T_{P1}$=330-370\,K and $T_{P0} \sim$ 620-650\,K. Both CDWs show a Fröhlich-mode transport. The LT-CDW formed below 150\,K also demonstrates sliding, but its nature is not fully understood.

The LT- CDW is observed only in samples belonging to the low-ohmic sub-phase and is not detected in the diffraction experiments. Both the amplitude of the feature in  $\sigma(T)$ at $T_{P2}$ and the density of the CDW current (at given $f_{f}$) depend strongly on the sample specific conductivity at RT, which can vary within the low-ohmic sub-phase by more than an order of magnitude. Evidently, the variations in specific conductivity are connected with the deficiency of S atoms. The excess free electrons, induced by "doping" from S vacancies, do not condense into a CDW until below 150\,K. The concentration of these electrons is comparable with the metallic value, so that the electronic gas can be considered as degenerate. Consequently, the electronic structure can be treated in terms of new Fermi surfaces, which survived the two upper CDW transitions. The 150\,K transition thus corresponds to condensation of these free electrons and, presumably, of additional electrons released during redistribution of S bonding, into a CDW, or a CDW-like formation. However, even for the low-ohmic samples the specific conductivity above $T_{P2}$ is far from being metallic. The free carriers are gapped by two already existing CDWs, and the resistivity is 2-3 orders of magnitude above the estimated value above $T_{P0}$ (Fig.\,8). Both, the $T_{P2}$ value and the form of the feature in  $\sigma(T)$ at $T_{P0}$ (Fig.\,14) show no obvious correlation with $\sigma _{s}$. Consequently, the transition at $T_{P2}$ is similar for the electron concentrations varying between samples by up to one and half orders of magnitude. In other words, the concentration of electrons condensed into the LT-CDW can vary a lot, but the characteristics of this CDW remain similar. The transition disappears or becomes invisible only if $\sigma _{s}$ is less than $\sim$ 10 ($\Omega$cm)$^{-1}$.

The possibility of forming a separate CDW by electrons from dopants has not been considered yet. In the cases of NbSe$_{3}$ and the monoclinic phase of TaS$_{3}$, where multiple Peierls transitions are also observed, different transitions dielectrize the electrons belonging to different types of chains. In the case of NbS$_{3}$-II, some electrons originate from the S vacancies acting as donors, and they are not expected to occupy a separate band. In case of the CDW compound K$_{0.3}$MoO$_{3}$ doping with V results in extra holes, but these are gapped by the same CDW~\cite{52}. Contrary to NbS$_{3}$-II, doping in this case only results in a variation of the $q$-vector.

No hysteresis was found in case of NbS$_{3}$-II in the $R(T)$, $R(V)$ and $R(\epsilon)$ measurements. Similar curves for nano-sized samples did not show steps coupled with the addition/removal of a CDW period (~\cite{7,53} and references therein). This could mean that the $q$-vectors of the two upper CDWs do not vary. NbS$_{3}$-II might have features in common with (TaSe$_{4}$)$_{2}$I, where the CDW also shows no metastable states (see~\cite{2}, pp. 357-360 and 364-367]). It seems that that the CDWs cannot deform in NbS$_{3}$-II like they do in K$_{0.3}$MoO$_{3}$, TaS$_{3}$ and a number of other compounds, probably because of topological reasons~\cite{54}. If this is the case, extra free electrons cannot be incorporated in the UHT- or RT-CDWs. Thus, "doping" with S-vacancies results in a growth of conductivity at $T$ $>$ $T_{P2}$, while at $T_{P2}$ the extra electrons become condensed into a separate CDW.

While the structure of NbS$_{3}$-I is well known~\cite{55}, the structure of NbS$_{3}$-II has not been determined yet. To analyze possible lattice instabilities of known phases of NbS$_{3}$ \emph{ab initio} density functional theory (DFT) calculations of a band structure were performed for a model structure consisting of symmetrized monoclinic unit cells of NbS$_{3}$-I. That is, we just manually removed the known dimerization of NbS$_{3}$-I and placed the atoms at the mean positions of a dimerized NbS$_{3}$-I unit cell~\cite{55}. This converted the compound into a metallic state. The DFT calculations were performed in both the local density approximations by the PAW (Projector Augmented-Wave) method~\cite{56}, as well as by the generalized gradient approximation~\cite{57}, as implemented in the Abinit simulation package~\cite{58}. Four bands were found to cross the Fermi level in the initial filling. Two of them have relatively large flat regions corresponding approximately to 1/2 and 2/3 filling of the respective bands. The corrugation of Fermi surfaces appears sensitive to the electron concentration: a reduced filling of these bands flattens the Fermi surfaces and makes the compound susceptible to CDW instability. These bands may be responsible for two transitions with two $q$-vectors. Two other bands form small near-cylindrical pockets around the Y point of the Brillouin zone and are aligned in the Y-H direction. They are more sensitive to doping and may eventually disappear in case of excessive doping. Extra electrons belonging to these pockets will not be condensed in the RT- or UHT-CDW, but may form a new condensed CDW state.

Several possibilities for the formation of the LT-CDW can be considered. The first is the Peierls transition. This would require electrons, not condensed by the CDW transitions at $T_{P1}$ and $T_{P0}$, to be in nested sections of the Fermi surface. This is not a very likely scenario, because of the strong dependence of the "fundamental ratio" on the electron concentration. There is a bigger probability that electrons belonging to these pockets will condense into a state in which the distance between the electrons depends on their concentration, both along and perpendicular to the chains. A candidate case in a Wigner crystal (WC), which is stabilized by the repulsive Coulomb forces. Unlike conventional CDWs, a WC is relatively weakly coupled with the lattice. This might be the reason, why the LT-CDW was not detected by TEM~\cite{59}.

However, $T_{P2}$, as well as the form of the $\sigma(T)$ curve in the logarithmic scale (Fig.\,14), remain stable over a wide range of electronic concentrations (nearly one and a half orders of magnitude). Since the temperature of a Wigner crystallization depends on the concentration of electrons as a power law, the observation argues against a WC formation at 150\,K.

The stability of $T_{P2}$ might indicate that the transition at 150\,K, forming a periodic potential, is not directly coupled with the electrons induced by doping. Nonetheless, these electrons, irrespective to their concentration, are accommodated into this potential and form a CDW electronic crystals. In case of the high-ohmic samples this condensation remains invisible in the $\sigma(T)$  curves. The transition at $T_{P2}$ can however show up in some other measurements like e.g. in  $\sigma(T)$ studies at sufficiently high frequencies (see Fig.\,4 in Ref.~\cite{15}). A kind of metal-dielectric transition was also observed close to 150\,K in the NbS$_{3}$-I polytype, where the transition temperature appears stable in a wide range of pressures~\cite{60}. Thus, the 150\,K transition seems to be an intrinsic feature of both NbS$_{3}$ polytypes and appears rather robust against pressure, strain and doping.

Apart from Wigner crystallization and Peierls transition, the Keldysh-Kopaev transition~\cite{23} (known also as the formation of excitonic dielectric~\cite{24}) has been suggested as a possible mechanism for electron condensation in NbS$_{3}$-II at $T_{P2}$~\cite{18}. This transition represents a generalization of the Peierls transition, which can occur if the electron's and hole's Fermi surfaces have shapes, which allow nesting. It may take place in a semiconductor, if the gap between the hole and electron bands is smaller than the binding energy of an exciton; then spontaneous exciton formation begins and a new electronic state develops. If the maximum and minimum of the hole and electron bands are displaced in the $k$-space, the vector connecting them defines the wave vector of the possible charge modulation, i.e. an excitonic CDW. Among related compounds such an origin of CDW formation has been suggested for TiSe$_{2}$~\cite{61}, which is also semiconducting above the transition. The nesting of electrons and holes with formation of excitons can proceed in a similar way for various degrees of doping. The dielectrization is pronounced if one of the bands (the electronic one in our case) is partly filled above $T_{P2}$. However, if the Fermi energy is located in the gap separating the bands, like it would be in case of a stoichiometric composition, one can expect a dielectric state already above $T_{P2}$ with the transition remaining practically invisible in $\sigma(T)$.

Neither the possibility of sliding nor the possible values of $j_{c}$/$f_{f}$ have been discussed for an excitonic CDW. One can consider a simple case of a "symmetric" excitonic CDW, which originates from the nesting of two similar electronic and hole bands, with a symmetrically positioned Fermi level ($p$=$n$). Such an excitonic CDW will not be charged. However, in case of doping, a sliding excitonic CDW can transfer charge proportional to the degree of doping.

\section{CONCLUSIONS.}

Our studies show that NbS$_{3}$-II is an outstanding member of the MX$_{3}$ group. It has three CDWs: a near room temperature RT-CDW at $T_{P1}$ = 360\,K, a low temperature LT-CDW at $T_{P2}$ = 150\,K, and another CDW at a much higher temperature $T_{P0}$  = 620-650\,K. Each CDW presents peculiar salient features. First, the RT-CDW has exceptionally high coherence and the utmost velocity of all known sliding CDWs. This suggests that this RT-CDW can be considered for practical applications. Second, the fields and currents of the UHT-CDW are impressive. Third, the LT-CDW is most unusual and a physical picture of its formation is still incomplete.

For readers' references we opt to summarize the features of NbS$_{3}$-II as a list:

\begin{enumerate}
  \item The properties of NbS$_{3}$- II depend strongly on the growth conditions. The RT conductivity,  $\sigma _{s}$(300 K), of the samples varies from 2 to 3$\times$10$^{2}$ ($\Omega$cm)$^{-1}$.
  \item The "low-ohmic" samples ( $\sigma _{s}$(300 K) = 10--3$\times$10$^{2}$ ($\Omega$cm)$^{-1}$ are S deficient with the S vacancies acting as electron donors. The gradual transformation of the "high-ohmic" into the "low-ohmic" ones under heating above $\sim$ 800\,K is consistent with this conclusion.
  \item The "high-ohmic" sub-phase ($\sigma _{s}$(300 K) = 2--10 ($\Omega$cm)$^{-1}$ of NbS$_{3}$-II shows two CDW transitions, at $T_{P1}$=340-370\,K and at $T_{P0} \approx$ 620-650\,K,
  \item Apart from the transitions at $T_{P0}$ and $T_{P1}$, the "low-ohmic" samples show a CDW transition at $T_{P2}$ = 150\,K. The specific conductivity drop at $T_{P2}$ is proportional to the specific conductivity above $T_{P2}$. This LT-CDW is a condensate of the electrons donated by the S vacancies.
  \item All three CDWs exhibit sliding at $E>E_{t}$.
  \item RF interference shows that the fundamental frequency of the RT-CDW sliding can be at least 20 GHz. The extremely high coherence of this CDW is manifested in Bessel-type oscillations of $E_{t}$ and of the Shapiro steps' width as a function of the RF power.
  \item At RT the NbS$_{3}$-II samples show torsional strain. The strain grows abruptly for $E > E_{t}$ and can be increased by an order of magnitude with RF irradiation.
  \item Under uniaxial stretching NbS$_{3}$-II samples the RT-CDW demonstrates many features of coherence enhancement, i.e., the threshold decreases and becomes sharper, the growth of  $\sigma _{d}$ above $E_{t}$ is faster, the value of the maximum CDW conductivity increases, and the Shapiro steps under RF irradiation become more pronounced.
  \item $T_{P1}$ is extremely sensitive to uniaxial stretching: $\epsilon \sim$ 1\% can reduce $T_{P1}$ to below RT. The transition at $T_{P1}$ becomes sharper with strain, in line with the growth of the CDW coherence.
  \item Sliding of the UHT-CDW can be observed below and above $T_{P1}$. This is proven by the effect of RF-induced coherence stimulation. The $E_{t}$ value for this CDW can be on the order of 10\,kV/cm at RT.
  \item The LT-CDW is nearly insensitive to tensile strain. At equal strain the relative decrease of $T_{P2}$ is 10 times lower than that of $T_{P1}$. The I-V curves show no regular changes under strain up to 1\%
  \item The charge density of the LT-CDW transport, i.e., the $j_{c}$/$f_{f}$  value revealed by RF interference, is from 3 to 1000 times smaller than that of the RT-CDW and scales with the specific conductivity above $T_{P2}$. The value $j_{c}$/$f_{f}$ appears to be well below 2$e$/s$_{0}$, which is impossible for a normal CDW.
  \item Unlike in cases of RT- and UHT- CDWs, no lattice distortion is observed at $T_{P2}$ by means of diffraction techniques. The $^{93}$Nb NMR study reveals a Knight shift at $T_{P1}$, but not at $T_{P2}$.
  \item A clear maximum near $T_{P2}$ is observed in the temperature dependence of the nuclear transverse magnetization relaxation rate, $T_{2}^{-1}$, measured at the central $^{93}$Nb NMR line. The feature in $T_{2}^{-1}(T)$ is interpreted by a freezing of the electronic density distortion (ordering) with cooling.
  \item Near $T_{P2}$, minima are observed in XANES spectra of the S K-edge, K-line pre-edge, and the Nb L3-edge. The result reveals electron transfer to both S 3$p$ and to Nb 4$d$ states down to $T_{P2}$. Further cooling below $T_{P2}$ reverses the trend of XANES intensity variation. A plausible cause for the observed XANES below $T_{P2}$ is a formation of excitons.
  \item The condensation of the excess electrons in the "low-ohmic" samples into a separate LT-CDW is explained with the rigidity of the RT- and UHT- CDWs, which does not allow changes of their $q$-vectors and condensation of extra electrons. In addition, \emph{ab initio} calculations show that these electrons may belong to additional small pockets in the Fermi surfaces, and a formation of a new condensed phase.
  \item The nature of the LT-CDW is not completely understood. The concept of an excitonic dielectric might explain the low sensitivity of $T_{P2}$ to the concentration of electrons, condensed into this CDW.
\end{enumerate}

Further studies of this compound, including structural and scanning-probe ones are in progress~\cite{22}. 

\begin{acknowledgments}
We are grateful to A.A. Sinchenko and A.P. Orlov for the help in the strain experiments. The support of RFBR (grants 14-02-01240, 14-02-92015, 16-02-01095) and the program 'New materials and structures' of RAS is acknowledged. The elaboration of the "bending technique" of the uniaxial expansion was supported by the Russian Scientific Foundation (Grant No14-19-01644). The work of I.R.M. was supported by the Program of Competitiveness Growth of Kazan Federal University funded by the Russian Government. The support by MOST, Taiwan (103-2923-M-002 -003 -MY3) and by the Slovenian Research Agency (ARRS) under the Slovenia-Russia bilateral project BI-RU/14-15-043 is also acknowledged. EPMA was performed, in part, using the equipment of MIPT Center of Collective Usage.
\end{acknowledgments}


\begin{thebibliography}{100}
\bibitem{20}	From this one can conclude that it is the $q_{1}$-CDW which provides the non-linear transport at RT.
\bibitem{21}	Evidently, in [11] the high-ohmic samples were reported, while in [13] - the low-ohmic ones (see below).
\bibitem{26}	Alternatively, the charge per chain transferred in time 1/$f_{f}$ can be 1e, as it was reported for the monoclinic phase of TaS$_{3}$~\cite{27}. This could be the case, if the CDW current in a cell is carried by 2 chains with opposite phases.
\bibitem{36}	This is seen from easy cleavage of the samples, from their high aspect ratio (length/width$\sim$10$^{5}$), from the elongated shape of the superlattice satellites [11]. The weak growth of the threshold field with cross-section reduction is also an indication of high anisotropy.
\bibitem{54}	In~\cite{11} it was noticed that within the experimental accuracy the $b$* component of $q_{1}$+2$q_{0}$ equals $b$*. It was suggested that $q_{1}$ could be the second harmonic of $q_{0}$. Alternatively, this relation could imply an extra condition for $q_{0}$ and $q_{1}$, forbidding their variation.
\bibitem{59}	Another probable reason for this is the large wavelength of the LT-CDW. To estimate the $q_{2}$-vector, we must first establish the electronic structure of NbS$_{3}$. 
The two upper $q$-vectors are close to 1/3$b$*. If we assume that at $T_{P0}$ two chains per unit cell are gapped, and at $T_{P1}$ - one (as found by the RF interference~\cite{19,12}),
the relation ($q_{1}$+2$q_{0}$)/3=1/3$b$* gives the value of the "average" $q$-vector and can be interpreted as an indication of 1/3 filling of the electronic band. Then the detailed 
formula of NbS$_{3}$ can be written as 2Nb$^{5+}$Nb$^{4+}$2(S$_{2}$)$^{2}$5S$^{2-}$. In this case there are 2/3 free electrons per Nb atom. Sulfur accepts either 1 or 2 electrons per atom. Then 5\% S vacancies can provide 0.15 to 0.3 electrons per Nb atom, and the correspondent CDW wavelength can range from 6.7 to 13.3 periods along the b axis. It is a good challenge to find such a distortion.
\end{thebibliography}
\end{document}